%% file: main.tex
\documentclass[iop,letter]{emulateapj}

\usepackage{natbib}
\setcitestyle{citesep={,}}
\usepackage{graphicx}
\usepackage{times}
\usepackage{apjfonts}
\usepackage{amsmath}
\usepackage{xcolor}
\usepackage{hyperref}
\hypersetup{    
  colorlinks      = true,
  linkcolor       = {black},
  linkbordercolor = {white},
  citecolor       = {black},
  citebordercolor = {white},
  urlcolor        = {blue},
  urlbordercolor  = {white},
}

\usepackage{lineno}
\input{commands}

\shorttitle{DES Processing Pipeline}
\shortauthors{Morganson et al.}

\begin{document}

\begin{nolinenumbers}
\vspace*{-\headsep}\vspace*{\headheight}
{\footnotesize \hfill FERMILAB-PUB-17-524-AE}\\
\vspace*{-\headsep}\vspace*{\headheight}
{\footnotesize \hfill DES-2017-0281}

\end{nolinenumbers}

\title{The Dark Energy Survey Image Processing Pipeline}

\input{authors}

\input{abstract}

\input{intro/intro}

\input{dataflow/dataflow}

\input{preprocessing/preprocess}
\input{preprocessing/cross}

\input{preprocessing/calibration}
\input{preprocessing/biascor}

\input{firstcut/firstcut}
\input{firstcut/astrometry}
\input{firstcut/initialmask}
\input{firstcut/skysub}
\input{firstcut/additionalmask}

\input{firstcut/psfex}
\input{firstcut/catalog}
\input{firstcut/eval}
\input{firstcut/finalcut}

\input{supernova/supernova}

\input{supernova/sncal}
\input{supernova/template}
\input{supernova/psfex}
\input{supernova/detection}

\input{supernova/dofake}
\input{supernova/caneval}
\input{supernova/sneval}

\input{coadd/coadd}
\input{coadd/coadd-tiles}
\input{coadd/coadd-astro}
\input{coadd/coadd-swarp}

\input{coadd/coadd-psfex}
\input{coadd/coadd-cat}
\input{coadd/coadd-post}


\input{summ/summ}
\input{acknowledgements}

\appendix

\input{appendix/flags}

\input{appendix/software1}

\input{appendix/software2}

\input{appendix/software3}

\end{document}

%% file: commands.tex
\usepackage{soul}
\usepackage{amsmath}
\usepackage{amssymb}
\usepackage{xspace}
\usepackage{xifthen}
\usepackage{eso-pic}

\newcommand{\maglimseg}{23.57}
\newcommand{\maglimser}{23.34}
\newcommand{\maglimsei}{22.78}
\newcommand{\maglimsez}{22.10}
\newcommand{\maglimsey}{20.69}

\newcommand{\FIXME}[1]{}

\newcommand{\COMMENT}[3]{}

\mathchardef\mhyphen="2D

\newlength{\dhatheight}

\newcommand{\bandvar}[2][]{%
  \ifthenelse{\isempty{#1}}{\var{#2}}{\var{#2\_#1}}%
}

\newcommand{\var}[1]{\ensuremath{\texttt{\MakeUppercase{#1}}}\xspace}

\providecommand\physrep{\ref@jnl{Phys.~Rep.}}%
\providecommand\apjs{\ref@jnl{ApJS}}%
\providecommand{\jcap}{\ref@jnl{JCAP}}%



%% file: authors.tex

\def\andname{}

\author{
E.~Morganson\altaffilmark{1,2},
R.~A.~Gruendl\altaffilmark{1,2},
F.~Menanteau\altaffilmark{1,2},
M.~Carrasco~Kind\altaffilmark{1,2},
Y.-C.~Chen\altaffilmark{1,2},
G.~Daues\altaffilmark{1},
A.~Drlica-Wagner\altaffilmark{3},
D.~N.~Friedel\altaffilmark{1,2},
M.~Gower\altaffilmark{1},
M.~W.~G.~Johnson\altaffilmark{1},
M.~D.~Johnson\altaffilmark{1},
R.~Kessler\altaffilmark{4},
F.~Paz-Chinch\'on\altaffilmark{1,2},
D.~Petravick\altaffilmark{1},
C.~Pond\altaffilmark{1},
B.~Yanny\altaffilmark{3},
S.~Allam\altaffilmark{3},
R.~Armstrong\altaffilmark{5},
W.~Barkhouse\altaffilmark{6},
K.~Bechtol\altaffilmark{7},
A.~Benoit-L{\'e}vy\altaffilmark{8,9,10},
G.~M.~Bernstein\altaffilmark{11},
E.~Bertin\altaffilmark{8,10},
E.~Buckley-Geer\altaffilmark{3},
R.~Covarrubias\altaffilmark{1},
S.~Desai\altaffilmark{12},
H.~T.~Diehl\altaffilmark{3},
D.~A.~Goldstein\altaffilmark{13,14},
D.~Gruen\altaffilmark{15,16},
T.~S.~Li\altaffilmark{3},
H.~Lin\altaffilmark{3},
J.~Marriner\altaffilmark{3},
J.~J.~Mohr\altaffilmark{17,18,19},
E.~Neilsen\altaffilmark{3},
C.-C.~Ngeow\altaffilmark{20},
K.~Paech\altaffilmark{17,21},
E.~S.~Rykoff\altaffilmark{15,16},
M.~Sako\altaffilmark{11},
I.~Sevilla-Noarbe\altaffilmark{22},
E.~Sheldon\altaffilmark{23},
F.~Sobreira\altaffilmark{24,25},
D.~L.~Tucker\altaffilmark{3},
W.~Wester\altaffilmark{3}
\\ \vspace{0.2cm} (DES Collaboration) \\
}

\affil{$^{1}$ National Center for Supercomputing Applications, University of Illinois at Urbana-Champaign, 1205 West Clark St., Urbana, IL 61801, USA}
\affil{$^{2}$ Department of Astronomy, University of Illinois at Urbana-Champaign, 1002 W. Green Street, Urbana, IL 61801, USA}
\affil{$^{3}$ Fermi National Accelerator Laboratory, P. O. Box 500, Batavia, IL 60510, USA}
\affil{$^{4}$ Kavli Institute for Cosmological Physics, University of Chicago, Chicago, IL 60637, USA}
\affil{$^{5}$ Department of Astrophysical Sciences, Princeton University, Peyton Hall, Princeton, NJ 08544, USA}
\affil{$^{6}$ University of North Dakota, Department of Physics and Astrophysics, Witmer Hall, Grand Forks, ND 58202, USA}
\affil{$^{7}$ LSST, 933 North Cherry Avenue, Tucson, AZ 85721, USA}
\affil{$^{8}$ CNRS, UMR 7095, Institut d'Astrophysique de Paris, F-75014, Paris, France}
\affil{$^{9}$ Department of Physics \& Astronomy, University College London, Gower Street, London, WC1E 6BT, UK}
\affil{$^{10}$ Sorbonne Universit\'es, UPMC Univ Paris 06, UMR 7095, Institut d'Astrophysique de Paris, F-75014, Paris, France}
\affil{$^{11}$ Department of Physics and Astronomy, University of Pennsylvania, Philadelphia, PA 19104, USA}
\affil{$^{12}$ Department of Physics, IIT Hyderabad, Kandi, Telangana 502285, India}
\affil{$^{13}$ Department of Astronomy, University of California, Berkeley,  501 Campbell Hall, Berkeley, CA 94720, USA}
\affil{$^{14}$ Lawrence Berkeley National Laboratory, 1 Cyclotron Road, Berkeley, CA 94720, USA}
\affil{$^{15}$ Kavli Institute for Particle Astrophysics \& Cosmology, P. O. Box 2450, Stanford University, Stanford, CA 94305, USA}
\affil{$^{16}$ SLAC National Accelerator Laboratory, Menlo Park, CA 94025, USA}
\affil{$^{17}$ Excellence Cluster Universe, Boltzmannstr.\ 2, 85748 Garching, Germany}
\affil{$^{18}$ Faculty of Physics, Ludwig-Maximilians-Universit\"at, Scheinerstr. 1, 81679 Munich, Germany}
\affil{$^{19}$ Max Planck Institute for Extraterrestrial Physics, Giessenbachstrasse, 85748 Garching, Germany}
\affil{$^{20}$ Graduate Institute of Astronomy, National Central University, Jhongli 32001, Taiwan}
\affil{$^{21}$ Universit\"ats-Sternwarte, Fakult\"at f\"ur Physik, Ludwig-Maximilians Universit\"at M\"unchen, Scheinerstr. 1, 81679 M\"unchen, Germany}
\affil{$^{22}$ Centro de Investigaciones Energ\'eticas, Medioambientales y Tecnol\'ogicas (CIEMAT), Madrid, Spain}
\affil{$^{23}$ Brookhaven National Laboratory, Bldg 510, Upton, NY 11973, USA}
\affil{$^{24}$ Instituto de F\'isica Gleb Wataghin, Universidade Estadual de Campinas, 13083-859, Campinas, SP, Brazil}
\affil{$^{25}$ Laborat\'orio Interinstitucional de e-Astronomia - LIneA, Rua Gal. Jos\'e Cristino 77, Rio de Janeiro, RJ - 20921-400, Brazil}

%% file: abstract.tex
\begin{abstract}

The Dark Energy Survey (DES) is a five-year optical imaging campaign with
the goal of understanding the origin of cosmic acceleration. DES
performs a $\sim 5000$ deg$^2$ survey of the southern sky in
five optical bands ($g,r,i,z,Y$) to a depth of $\sim$24th magnitude.
Contemporaneously, DES performs a deep, time-domain survey in four
optical bands ($g,r,i,z$) over $\sim$ 27 deg$^2$.  DES exposures
are processed nightly with an evolving data reduction pipeline and
evaluated for image quality to determine if they need to be retaken.
Difference imaging and transient source detection are also performed
in the time domain component nightly.  On a bi-annual basis, DES exposures are reprocessed with a refined pipeline and coadded to maximize imaging depth.  Here we
describe the DES image processing pipeline in support of DES science,
as a reference for users of archival DES data, and as a guide for
future astronomical surveys.

\end{abstract}


%
%
%

%% file: intro/intro.tex
\section{Introduction}\label{sect:intro}

The Dark Energy Survey (DES) is a large optical imaging survey designed 
to improve our understanding of the origin of cosmic acceleration and the nature of 
dark energy using four complementary methods: weak gravitational lensing, 
galaxy cluster counts, the large-scale clustering of galaxy (including baryon 
acoustic oscillations), and distances to Type Ia supernova \citep{2005astro.ph.10346T}. 
DES uses the $3 \deg^2$ Dark Energy Camera \citep[DECam;][]{2015AJ....150..150F}, 
a 570 Megapixel camera installed at prime focus on the Blanco 4-m
telescope at the Cerro Tololo Inter-American Observatory (CTIO) in northern 
Chile. Each exposure travels from CTIO to the National Optical Astronomical Observatory (NOAO) for archiving 
and then to the National Center for
Supercomputing Applications (NCSA) at the University of Illinois at 
Urbana-Champaign for processing within minutes of being observed.

DES began in August 2013 and observes for roughly six months out of the year (August through mid-February). DECam Science Verification (SV) data were taken between 2012 November 1 and 2013 February 22 were of sufficient quality to be included in some DES data sets. 
Raw exposures are publicly available from NOAO one year after acquisition, and DES is scheduled to provide two public releases of processed data.\footnote{\url{https://des.ncsa.illinois.edu/releases}}

\begin{figure}
\centerline{
\includegraphics[width=\columnwidth]{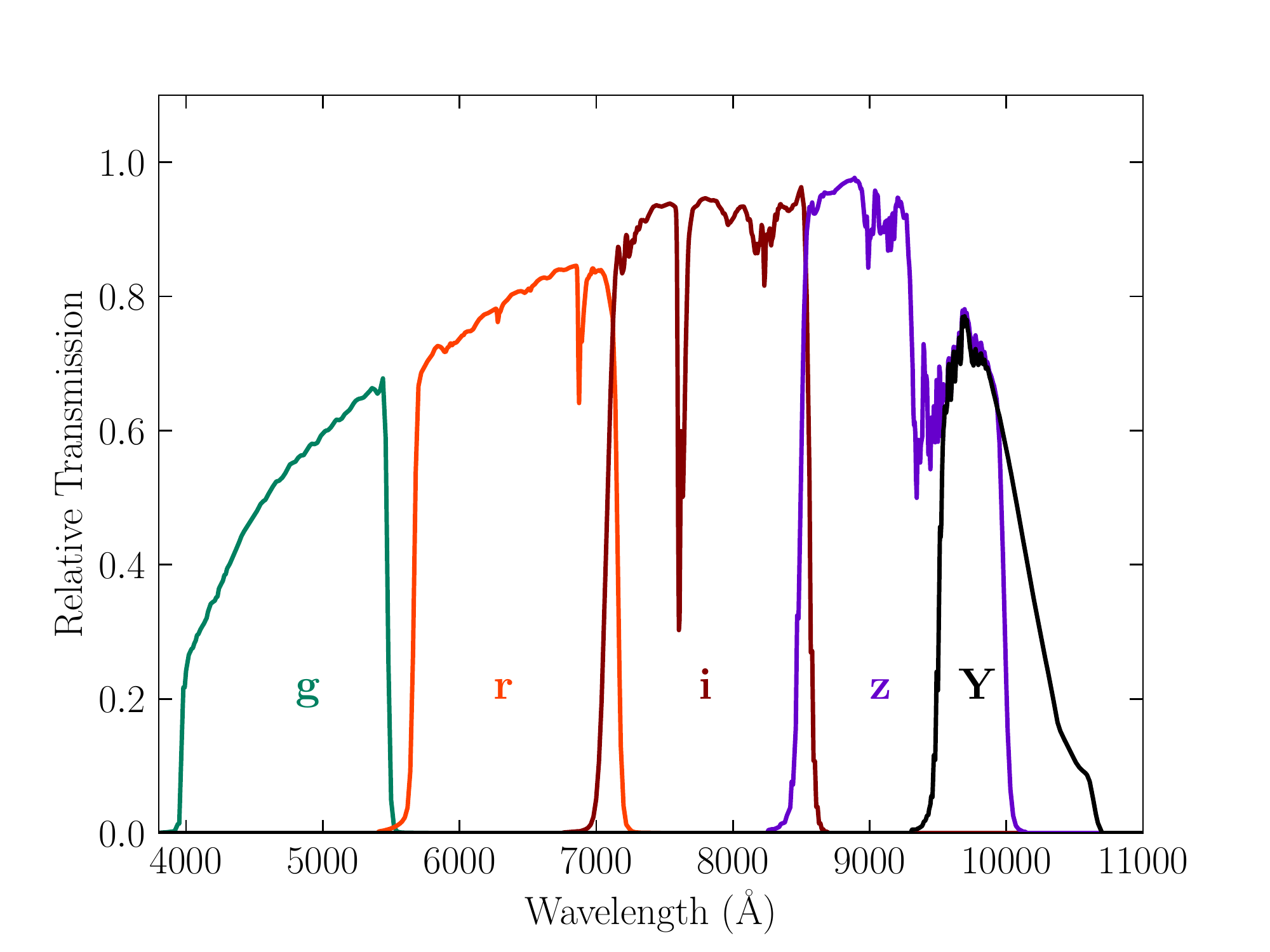}
}
\caption{The DES $g$, $r$, $i$, $z$ and $Y$ Standard Bandpasses. These curves include atmospheric transmission and instrumental response. Details can be found in \citet{FGCM}.}\label{fig:filters}
\end{figure}

DES is one in a series of large optical surveys that includes the Sloan
Digital Sky Survey \citep[SDSS;][]{2000AJ....120.1579Y}, the Panoramic
Survey Telescope and Rapid Response System 1 \citep[Pan-STARRS1 or
  PS1;][]{2010SPIE.7733E..0EK}, Kilo Degree Survey \citep[KiDS;][]{2013ExA....35...25D}, 
the Hyper Suprime-Cam SSP Survey \citep[HSC;][]{2017arXiv170405858A} and the 
future Large Survey Synoptic Telescope \citep[LSST;][]{2008arXiv0805.2366I}. DES 
comprises two multi-band imaging surveys: a
$\sim 5000 \deg^2$ wide-area survey of the southern sky in the $g$, $r$, $i$, $z$ and $Y$ bands and a survey of $\sim 27 \deg^2$ observed
with high cadence in the $g$, $r$, $i$ and $z$ bands to detect
supernovae and characterize their light curves. The DES filters are very
similar to their analogously-named counterparts from other surveys. 
Uniquely, the DES $z$ band extends redward of, for instance, the SDSS $z$ band 
and largely overlaps with the DES $Y$ band (Figure \ref{fig:filters}). DECam also has a near ultraviolet $u$ filter, a wide optical $VR$ filter, and a narrow $N964$ filter. None of these are used directly 
by DES, although some supplemental $u$-band data has been taken in the DES supernova fields 
and elsewhere. 

\begin{figure}
\centerline{
\includegraphics[width=\columnwidth]{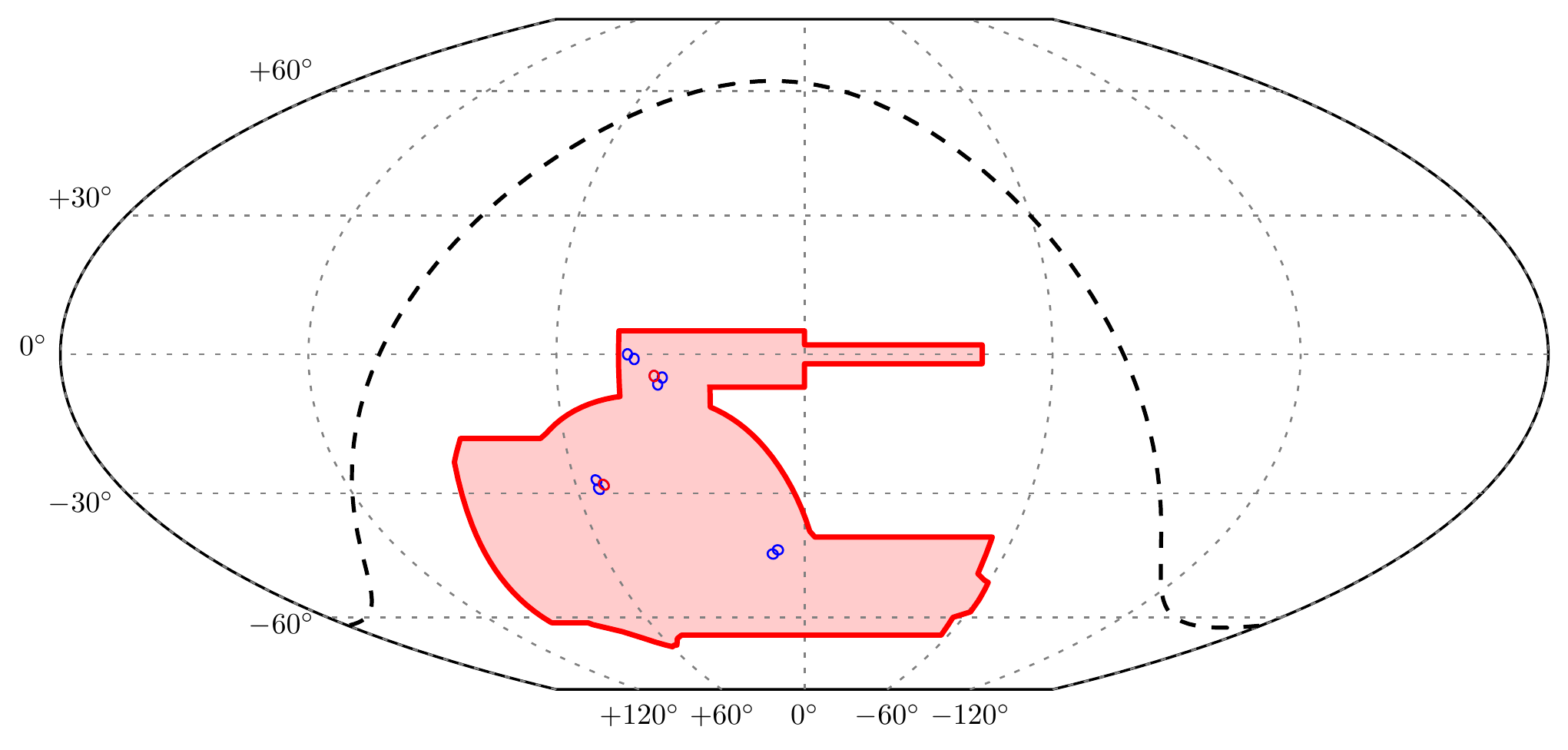}
}
\caption{An equal-area McBryde-Thomas flat-polar quartic projection of the DES footprint in celestial equatorial coordinates. The main 5000 deg$^2$ survey footprint for DES is shown in red. The 8 shallow supernova fields are shown as blue circles, and the 2 deep supernova fields are shown as red circles. The Milky Way plane is shown as a dashed line.}\label{fig:footprint}
\end{figure}

The DES wide-area survey footprint covers $\sim 5000\deg^2$ of the southern Galactic cap (Fig. \ref{fig:footprint}), overlapping SDSS Stripe 82 along the celestial equator and $\sim 2500 \deg^2$ of South Pole Telescope footprint \citep{2011PASP..123..568C}.
The DES wide-area survey images have a nominal $10\sigma$ limiting magnitudes of $g = \maglimseg$, $r = \maglimser$, $i = \maglimsei$, $z = \maglimsez$ and $Y = \maglimsey$.\footnote{Limiting magnitudes were calculated from the median PSF magnitude of all sources with ${\rm S/N} = 10$ detected in exposures taken under nominal observing conditions. These estimates of the limiting magnitudes have an RMS spread of roughly $\pm 0.09$ mag.}
The final coadded depth will be roughly one magnitude deeper. DES wide-area images have a median seeing of $\sim$0\farcs99.

To accomplish the DES scientific goals, the DES Data Management system
\citep[DESDM, ][]{2006SPIE.6270E..23N, 2008SPIE.7016E..0LM, 2011arXiv1109.6741S, 2012SPIE.8451E..0DM, 2012ApJ...757...83D} 
generates a variety of scientific products. For the DES wide-area survey, 
DESDM produces reduced single-epoch images, deeper coadded images, and catalogs of suitable quality to perform photometric redshift, weak lensing, and other precise analyses. 
DESDM  produces difference images and catalogs for each observation of a 
supernova field. In addition to facilitating the cosmological analyses that 
primarily motivated DES, these data products will provide a rich legacy for 
the astronomical community \citep{2016MNRAS.460.1270D}.

The processing pipeline described here was first applied to a cumulative processing of data from Year 1 (Y1) through Year 3 (Y3), and serves as the foundation of the first DES public data release \citep[DES DR1;][]{DR1}.
This pipeline differs somewhat from that used to create the Y1A1 data products \citet{Y1A1} and the associated Y1A1 GOLD catalog used for DES Y1 cosmology results (e.g., \citealt{Y1KP} and associated papers). 
We attempt to describe places where the two pipelines differ appreciably. 

We document here how the DES data products, processed images and catalogs, 
are generated from raw DECam images. In the next section, 
we summarize the flow of data from the camera through our processing 
pipelines. In Section \ref{sect:preprocess}, we describe the preprocessing: 
generating image calibration data and handling camera-based artifacts. 
Section \ref{sect:firstcut} describes the First Cut pipeline, the nightly 
processing of wide-area data primarily used for data assessment, and the minor 
changes we made for our ``Final Cut'' processing for public release. Section 
\ref{sect:supernova} describes our nightly supernova processing. 
We describe our Multi-Epoch pipeline, which includes coaddition of images and more precise photometry,
in Section \ref{sect:coadd}. 

%% file: dataflow/dataflow.tex
\section{Data Flow}

\begin{figure*}
\centerline{
\includegraphics[width=7.3in]{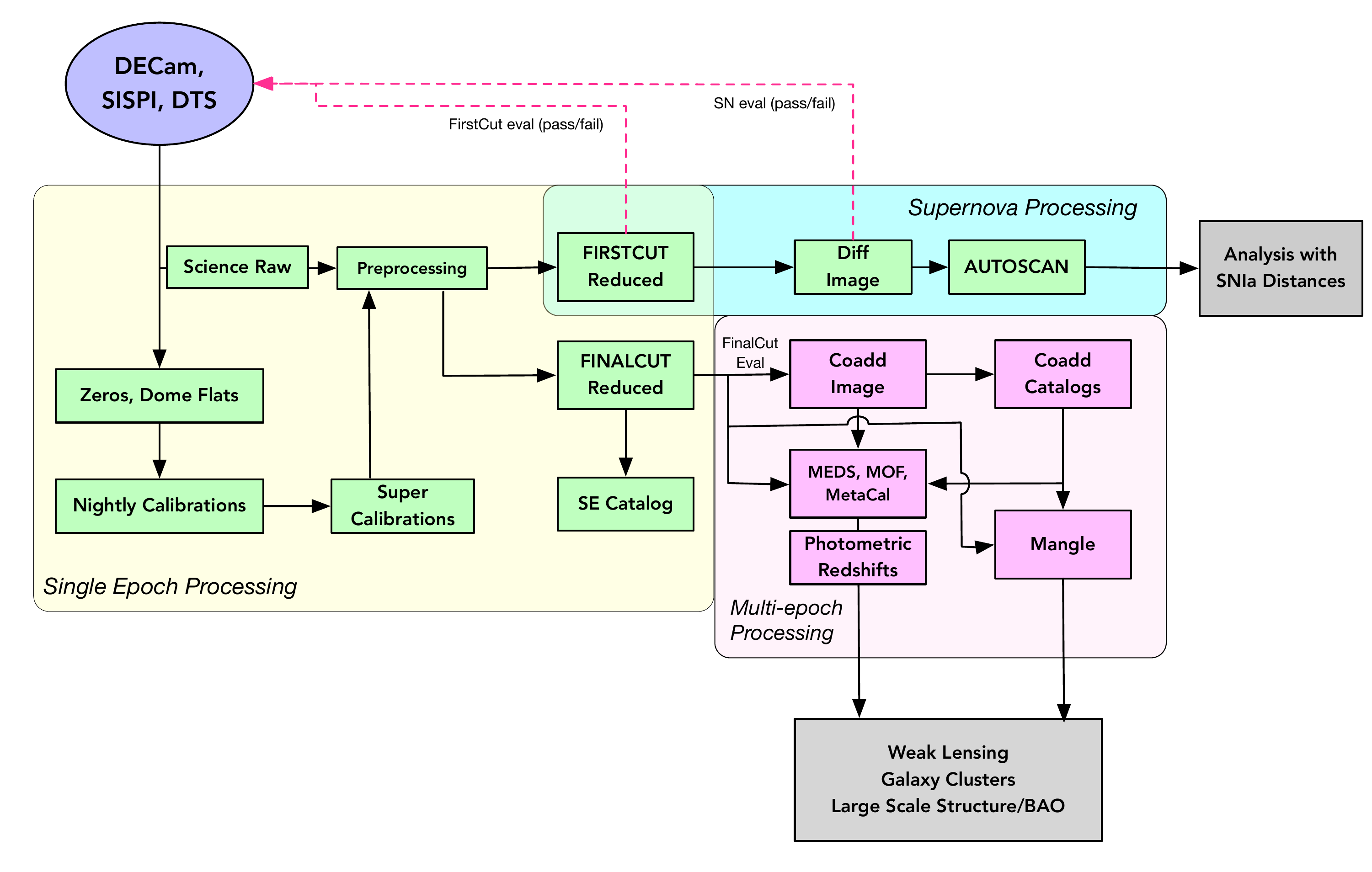}}
\caption{A schematic view of the modules and steps involved in the
  DESDM Pipeline. Calibration and raw science images are delivered from DECam by SISPI and DTS to NCSA. In Single Epoch Processing (yellow box), raw science exposures undergo preprocessing and are combined with calibration 
images to produce First Cut images for initial inspection and Final Cut images for science releases. Images in the Supernova fields are processed with the Supernova Pipeline (blue boxes). Final Cut images and catalogs are 
sent to Multi-Epoch Processing (pink box), the outputs of which go to the science groups for validation and analysis (gray boxes). The science pipelines are not described in this paper.}\label{fig:sispi}
\end{figure*}

In this section we provide a high-level overview of the DES data flow starting with mountain-top observations and concluding with final processed data products (Fig.\ \ref{fig:sispi}).
DES observations are scheduled by the Observer Tactician software \citep[\textsc{ObsTac;}][]{ObsTac}, which determines the on-sky observing strategy
based on observing conditions and current survey needs using data quality assessments from previously processed 
data. Our 10 supernova fields are observed every seven days or when the seeing exceeds 1\farcs1. 
After each observation,  the Survey Image System Process Integration \citep[SISPI;][]{2012SPIE.8451E..12H} transfers 
the raw exposures to the NOAO via the Data Transport 
System \citep[DTS;][]{2010ASPC..434..260F} for archiving. 
Exposures continue from NOAO to the National 
Center for Supercomputing Applications (NCSA) for processing. Calibration 
images (bias, dark, and dome flat exposures) are delivered nightly and processed into the format used for flat fielding (Section \ref{sect:calibration}). 

\begin{figure}
\centerline{
\includegraphics[width=3.3in]{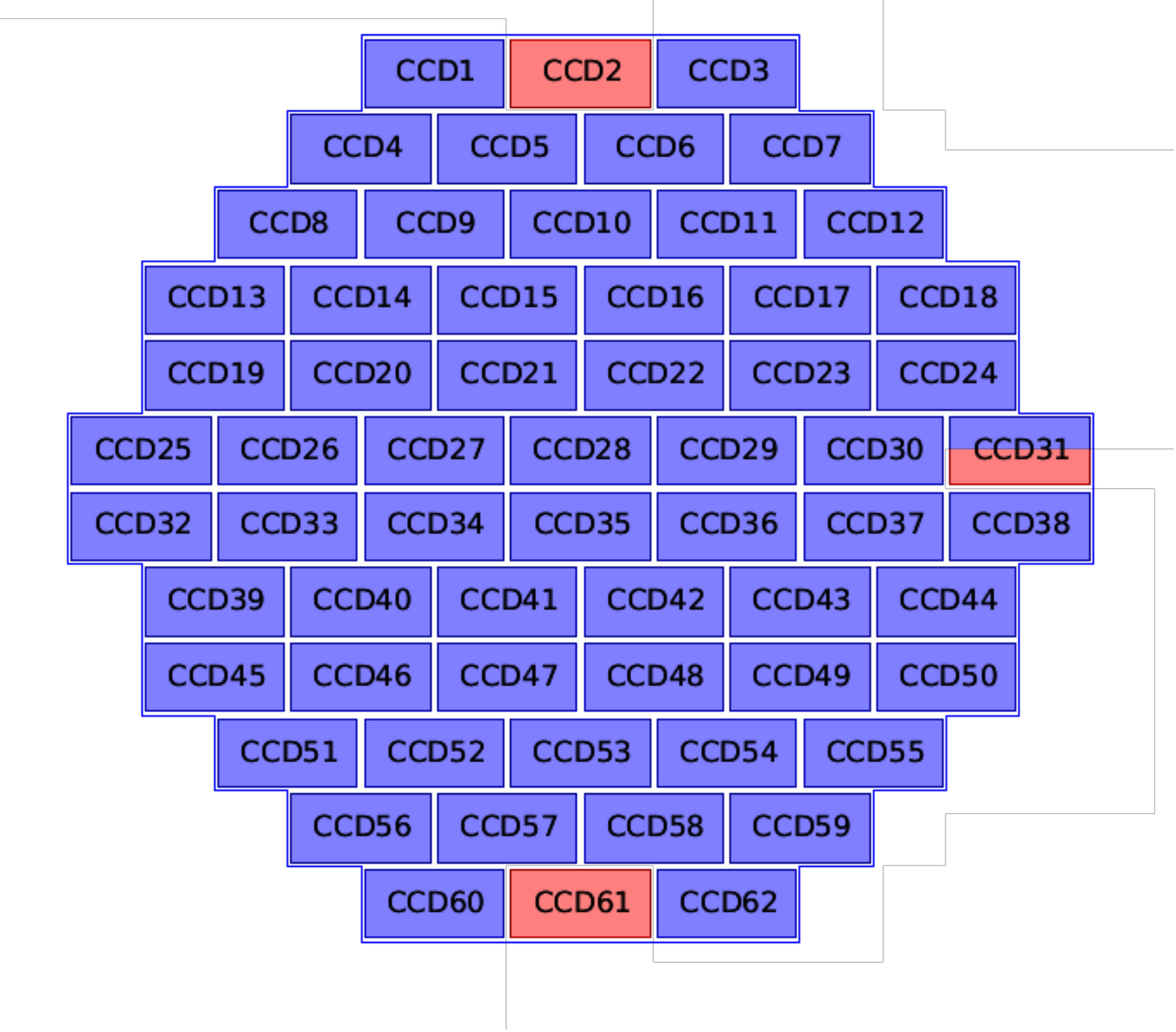}
}
\caption{DECam focal plain chips layout. Each of the 62 rectangles represents a CCD which is in 
turn read by two amplifiers (along the long direction). CCDs 2 and 61 have been inactive for most of DES. Amplifier A of CCD 31 has unpredictable gain and has not been processed by DES. These areas are marked in red.}\label{fig:DECam_layout}
\end{figure}

The DES data are collected by DECam, a 570 Megapixel camera at the Blanco 4-m telescope at the CTIO in Chile. 
The DECam focal plane is comprised of 62 science CCDs (see Fig. \ref{fig:DECam_layout}) which are each readout by two amplifiers. 
DECam has 8 additional corner CCDs for guiding, focusing and alignment. 
These 8 CCDs are not reduced or analyzed in our pipelines. In addition, the routines described here are not used to process CCD 61, which failed in 2012. 
Similarly, CCD 2 was not processed between November 2013 and December 2016, during which time it was not functional.
Amplifier A of CCD 31 has an unpredictable, time-variable gain and is not processed or used for DES science. 
Amplifier B of CCD 31 functions normally and is processed.
The rest of the science CCDs are performing within specifications and are usable for science. 
Thus, we have used either 59.5 or 60.5 CCDs throughout most of the survey. 
Each science CCD has 2048 $\times$ 4096 pixels and the pixel scale varies smoothly from 0\farcs2636 per pixel down to 0\farcs2626 per pixel as one travels from the center of the camera focal plane out to its edge, 1.1 degrees away. 
The total active area covered by all science CCDs is roughly $2.7 \deg^2$. 
Each DECam amplifier is readout with a virtual overscan comprising 6 pre-scan pixels at the beginning of each row and 50 overscan pixels at the end of each row.
The DECam CCDs are readout at a pixel rate of 250 kHz (per amplifier) setting an overall readout time of 17 seconds.
This readout time is less than the typical overhead and slewing time between exposures.

DES has both a wide-area and a supernova component. These different
survey components require different exposure times, described in
Table \ref{tab:exptime}. In the wide-area survey, exposures are 90 seconds 
in the $g$, $r$, $i$ and $z$ bands, and the 5-year survey goal is to obtain 10 
exposures in each band. Initially, DES planned to obtain 
ten 45 second exposures in the $Y$ band, but in Y4 DES began collecting one 90 
second exposure per year to reduce overhead. The total exposure time should be 900 seconds in the 
$g$, $r$, $i$ and $z$ bands and 450 seconds in the $Y$ band. To optimize for 
supernova detection, the ten supernova fields are observed to the same limiting magnitude depth 
in the $g$, $r$, $i$ and $z$ bands. This optimization requires longer exposure 
time in the redder bands as described in Table \ref{tab:exptime}. We will ultimately visit each supernova field 120 times, providing total exposure times much 
longer than the wide-area survey.

\begin{table}
\centering
\begin{tabular}{cccc}
        \hline
       &           & Shallow   & Deep      \\
Filter & Wide field & Supernova & Supernova \\
        \hline
$g$    & 90        & 175 & 200 $\times$ 3\\
$r$    & 90        & 150 & 400 $\times$ 3\\
$i$    & 90        & 200 & 360 $\times$ 5\\
$z$    & 90        & 200 $\times$ 2 & 330 $\times$ 11 \\
$Y$    & 45/90     & -- & -- \\
        \hline
\end{tabular}
\caption{Exposure times (in seconds) for DES wide-area and supernova exposures. Deep Supernova fields and the Shallow Supernova field $z$-band images are constructed of multiple exposure ``coadds'', and we note the number of exposures as appropriate. In Y4, the $Y$-band exposure time went from 45 seconds to 90 seconds.}\label{tab:exptime}
\end{table}

DES wide-area data are processed with the First Cut pipeline (Section \ref{sect:firstcut}).
Wide-area exposures are typically processed within 10 hours of being taken, which allows for rapid evaluation of data quality (Section \ref{sect:eval}).
The supernova pipeline (Section \ref{sect:supernova}) begins by processing images 
identically to the First Cut pipeline. It then produces a difference image with the nightly image and a previously generated 
template to identify transient events. 
The supernova data are given higher priority in our processing queue to ensure that they are processed and evaluated before the subsequent night.
The goal of the First Cut pipeline is to determine whether or not an exposure meets
survey quality requirements for image quality and image depth (Section \ref{sect:eval}). 
Supernova processing has an analogous image evaluation procedure (Section \ref{sect:sneval}). 
If an exposure does not satisfy the data quality criteria for the relevant survey, then that field-band combination is placed back on the stack for observation.
Data evaluations are usually completed before the start of observing the following night.

The single epoch images are processed multiple times. The aforementioned First Cut pipeline is run immediately for image evaluation. 
Final Cut processing occurs at the end of a season, typically in March, and is a full reprocessing of all the single epoch
exposure from the prior 6 months.
The Final Cut processing uses carefully constructed calibration images for flat fielding and sky characterization (see
Section \ref{sect:finalcut}) and a single pipeline that includes all improvements developed during the observing season. 
A rough photometric calibration ($\approx$ 5\% RMS scatter) is performed by the pipelines above. More precise zeropoints are assigned to the 
Final Cut data with the calibration procedure described in \citet{FGCM}. 

After Final Cut, we employ a Multi-Epoch pipeline to maximize survey
depth by coadding images into a single wide-area image (Section
\ref{sect:coadd}). We create a catalog of astronomical objects from 
these images. We also produce depth maps which are used to model
survey depth as a function of position on the sky using \textsc{mangle}
\citep{2004MNRAS.349..115H,2008MNRAS.387.1391S} and produce Multi
Epoch Data Structure \citep[MEDS;][]{2016MNRAS.460.2245J} postage stamps of every object. We then perform improved photometric and morphological measurement with multi-epoch, multi-object fitting \citep[MOF;][]{Y1A1}, and 
Metacalibration \citep[Metacal;][]{2017arXiv170202601S}, a technique to accurately measure weak gravitational lensing shear.
After source detection and measurement,
the DES data are ready to be analyzed by high-level science algorithms.

Over the first four years of the survey (i.e. as of 2017 June 15), DES has taken 67130 wide-area exposures and 9571 
supernova exposures. We anticipate $\sim$85000 and $\sim$12000 wide-area and supernova exposures when the 
survey is complete. This excludes photometric calibration and astrometric alignment images. 
The average raw exposure (compressed losslessly by a factor of roughly 2) is 0.47 GB, and we have 
processed 36 TB of raw science data. This expands to 128 TB if we include calibration images and other 
ancillary data, and balloons to 1.0 PB of processed image and catalog products. The wide-area survey 
and supernova images comprise 1510 hours and 680 hours of DECam time, respectively. Each 
reprocessing of the wide-area data requires roughly 500,000 core hours. The more 
intensive supernova processing, which includes difference imaging, requires 600,000 core hours. In addition, 
DESDM used 200,000 core hours to process 10,346 coadded tiles. Including all reprocessing campaigns, DESDM 
has used roughly 2 million core hours at a combination of the test cluster at NERSC, NCSA's iForge cluster, Fermilab's Fermigrid batch computing 
system \citep{2008JPhCS.119e2010C}, the University of Illinois campus cluster and the Blue 
Waters supercomputer \citep{BlueWaters}. We use Oracle Real Application Cluster 12c to support development, 
testing, nightly operations, data release production, and data distribution. The total 
size of the processing database (containing processing log information) and the science release database (containing objects catalogs) are currently 36 TB and 17 TB, 
respectively.

%% file: preprocessing/preprocess.tex
\section{Preprocessing}\label{sect:preprocess}

\begin{figure*}
\centerline{
\includegraphics[width=7.3in]{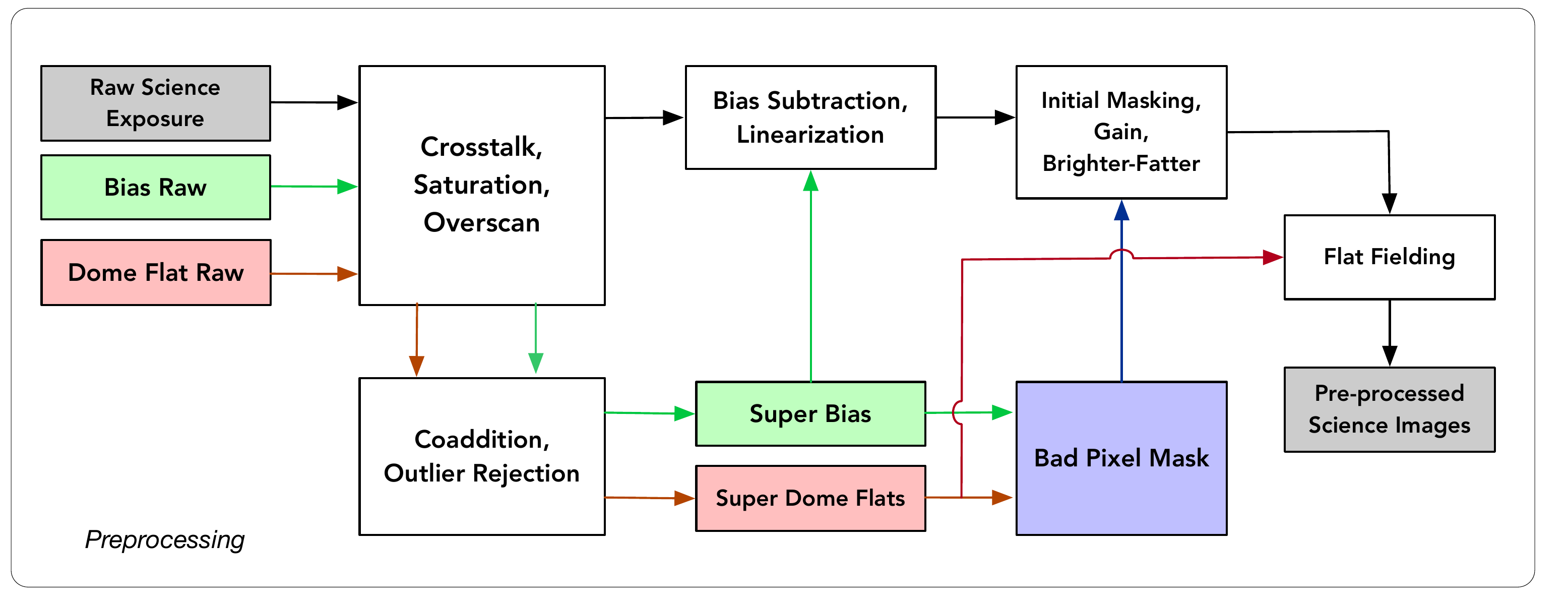}
}
\caption{A diagram of the Preprocessing applied to every exposure. This is a ``zoom in'' of the the Preprocessing box in Fig.\ \ref{fig:sispi}. This includes our initial masking and detrending with bias and flat images (green and red arrows, respectively). 
Preprocessed exposures are then processed with the First Cut or Final Cut pipeline.}\label{fig:preprocess}
\end{figure*}

All raw DECam exposures require a set of generic corrections  
before image-specific processing. We call these routines ``Preprocessing'' and
diagram them in Fig.\ \ref{fig:preprocess}, a zoom in of the Preprocessing box from Fig.\ \ref{fig:sispi}. 
These procedures include crosstalk correction, bias subtraction, bad pixel masking 
(masking known problematic pixels in the camera), correcting nonlinear pixel 
response, and flat fielding. These corrections in turn require a series of 
engineering tables for crosstalk and linearization, as well as 
bias images, dome flats, star flats, bad pixel masks and sky background model images. We 
describe our preprocessing routines, tables and images below. 
Additional flat-fielding routines are performed later 
in the processing in Section \ref{sect:skysub}. Many of the 
underlying phenomena here have their mathematical and physical basis described in 
detail in \citet{Bernstein2017}. 

Preprocessing requires many auxiliary files and tables which we describe Table \ref{tab:files} and Table \ref{tab:links} of Appendix \ref{sect:flags}.  
Table \ref{tab:files} lists, for each auxiliary correction element, the element's 
name, a brief description, how often the element is updated and which code module applies the element to the data during 
processing.
Table \ref{tab:links} documents the bias and dome flat correction images, and also
special tables and images to correct for other features such as non-linear CCD response. 
These auxiliary correction files and tables are referenced in Sections \ref{sect:calibration}-
\ref{sect:skysub} below.  Code to apply the correction images is publicly available.\footnote{\url{https://github.com/DarkEnergySurvey}}

A DES ``processing epoch'' is declared at the beginning of each observing year, and also when an event, such as a camera 
warm-up/cool-down cycle, could change the characteristics of the camera response (requiring new flat fields) 
or pixel-defect maps (new bad pixels appearing).  Events that cause a significant change to the camera temperature 
can change the astrometry (relative orientation of the CCDs within the camera) by tiny amounts (a few microns), 
and can also change the flat field and other response of the camera by small amounts ($\ll 1\%$, but large enough to 
warrant generating a new set of auxiliary correction files). A list of processing epochs is shown in Table \ref{tab:epochs}.
Also, of note, baffling was added to reduce scattered light 
for the $g$ band on 2013 April 22 and the other bands on 2013 August 12. 
The internal part of shutter and filter 
units were aerosolized (painted) on 2014 March 14.

\begin{table}
\centering
\begin{tabular}{cccc}
        \hline
Epoch Name & Begin Date & End Date \\
        \hline
SVE1 & 2012.11.09 & 2013.02.22\\
Y1E1 & 2013.08.15 & 2013.11.28\\
Y1E2 & 2013.11.30 & 2014.02.12\\
Y2E1 & 2014.08.06 & 2014.11.30\\
Y2E2 & 2014.12.04 & 2015.05.18\\
Y3E1 & 2015.07.31 & 2016.02.23\\
Y4E1 & 2016.08.13 & 2016.12.31\\
Y4E2 & 2017.01.01 & 2017.02.18\\
Y5E1 & 2017.08.15 & 2017.09.04\\
Y5E2 & 2017.09.04 & --\\
        \hline
\end{tabular}
\caption{\rm{A list of epoch names, begin dates and end dates. Date are in year.month.day format. The epochs are named for the survey year in which they were taken (with Science Verification data called SV) and then the epoch number within that year. So Y4E2 is the second epoch from the fourth year of the survey.}}\label{tab:epochs}
\end{table}

%% file: preprocessing/cross.tex
\subsection{DECam Crosstalk}\label{sect:crosstalk}

Crosstalk occurs when the signal in one amplifier contaminates another. 
A crosstalk correction is generally necessary for any camera with multiple 
CCDs or amplifiers \citep{2001ExA....12..147F}. DES has measurable crosstalk 
between the two amplifiers on each CCD and sometimes between amplifiers on different CCDs. 
We correct for this crosstalk in both science exposures and all calibration 
images (Section \ref{sect:calibration}). 

DECam crosstalk was evaluated by comparing a series of images with 
bright stars randomly located across the focal plane. 
We correlate the flux that is measured by each ``Source'' amplifier on CCDs where the stars are imaged, to the flux measured for every other amplifier on the camera.
Each other amplifier is a potential ``Victim'', and we check for correlated charge measured in the pixel location that is read out simultaneously with the bright star.
 We fit for a linear coefficient to correct every Victim amplifier for ``Source-Victim'' combination:

\begin{equation}
\Delta N_{Victim} = -C_{VS} N_{Source}.
\end{equation}\label{eq:pixcorrect}

Here $N_{Victim}$ and $N_{Source}$ are the fluxes in the corresponding 
pixels of the Victim and Source CCD-amplifiers. $\Delta N_{Victim}$ is the flux 
above background observed in the Victim amplifier. $C_{VS}$ is the fitted 
coefficient and is different for every CCD-amplifier pair (see link in Appendix 
\ref{sect:flags}, Table \ref{tab:links} for $C_{VS}$ values). Absolute $C_{VS}$ values measured
to be less than $10^{-6}$ are set to 0. $C_{VS}$ values between amplifiers 
on the same CCD can be as high as $10^{-3}$. All $C_{VS}$ values between CCDs are measured 
to be less than $2 \times 10^{-4}$ and most are less than $10^{-5}$ 
\citep{Bernstein2017}. These coefficients constitute a Crosstalk Matrix described in Appendix
\ref{sect:flags}, Table \ref{tab:files}. They were calculated at the beginning of the survey, and have 
been left unchanged as no significant change has been detected.

The routine that performs the crosstalk correction also runs an overscan correction to remove 
instrumental bias. Each amplifier has a 50 column wide unilluminated overscan region that spans 
its entire length. For each row, we compute the mean number of counts across the middle 
40 pixels of the overscan region (after removing the minimum and maximum pixel 
value) and subtract it from the entire row. The overscan mean is calculated
independently for each row with no fitting or interpolation between neighboring rows. The use of a row-by-row 
overscan is particularly important for DECam, because CCDs that share a backplane with the 
focus CCDs see a bias jump when the focus CCDs finish reading halfway through the science CCD
readout.  The crosstalk code also splits the full raw exposure into separate single-CCD images, 
and the single raw Analog-Digital Unit (ADU) plane is converted to a three-layer multi-HDU fits image, consisting of 
flux, mask and weight layers as HDUs 0, 1 and 2. 


%% file: preprocessing/calibration.tex
\subsection{Generating Calibration Images}\label{sect:calibration}

To produce images of sufficient quality for cosmological analysis, we account for a variety 
of instrumental effects using a series of calibration images. We first perform initial flat fielding  
of exposures using bias and dome flat corrections.  We also correct for CCD nonlinearity and the 
brighter-fatter effect \citep{Antilogus2014,2015JInst..10C5032G}. We produce a Bad Pixel Mask (BPM) to record DECam-specific 
artifacts as well as sky background templates and star flats to facilitate sky background subtraction from 
each science image. We remake the above calibration files each time the camera warms up, typically twice 
per season. 

Bias and dome flat images for each filter are taken nightly. 
In the first year of data taking, these nightly calibration
images were used to detrend science exposures during nightly processing. 
However, it was found that the camera response was more stable than the individual nightly calibration exposures, and after the first year we used a set of ``supercals'' assembled by combining flats and biases taken over several nights.
Our bias images are composed of a mean coadd with $5\sigma$ rejection of 
approximately 150 null exposure time images that have been crosstalk corrected. 
Similarly, the flat field images are $5\sigma$ rejected mean coadds of typically 
50-100 short (8s-20s, depending on filter) illuminated dome-flat
images that have been crosstalk corrected. 
A single set of these averaged supercals are adequate for removing the pixel-to-pixel response of the camera for an observing epoch.
Nightly calibration images are still used to evaluate camera function.

We also correct for CCD nonlinearity and the so-called ``brighter-fatter'' effect
in all science images and flat images. 
\citet{2010SPIE.7735E..1RE} report DECam CCD nonlinearity at both low fluxes 
and near saturation using flat field images. We correct for this nonlinearity 
with a CCD-dependent lookup table that converts observed flux to their fitted 
model of true flux \citep{Bernstein2017}. The brighter-fatter effect is caused by 
charge building up on well-illuminated pixels and pushing away nearby charge, effectively broadening 
the PSF of bright objects. \citet{2015JInst..10C5032G} model this effect for 
DECam, and we use their CCD-dependent kernel to account for the 
brighter-fatter effect in the data.

Associated with each observing epoch, we produce a BPM from the flats and biases
along with a set of approximately 50 $g$-band sky exposures. The BPMs 
are propagated into the mask plane for each CCD image, which also includes 
``bad pixel'' information specific to each exposure. The BPM bit mask is provided 
in Appendix \ref{sect:flags}, Table \ref{tab:bpm}. Pixels with values less than 50\% 
the mean (over the full camera) in any of the inputs to the flat coadd image are flagged as being cold 
(\texttt{FLAT\_MIN}). Similarly, pixels with values greater than 150\% of the mean 
in three or more input images are marked as hot (\texttt{FLAT\_MAX}). 
Pixels with more than 500 photo-electrons in any of the input bias images are flagged as 
hot (\texttt{BIAS\_HOT}) while those with more than 20 photo-electrons are flagged 
as warm (\texttt{BIAS\_WARM}). Pixels within a $11 \times 3$ rectangle (along the readout 
direction) centered at a hot pixel are also marked as warm. Columns with multiple hot pixels 
are marked as \texttt{BIAS\_COL}. All pixels read out after a hot pixel are also
marked as \texttt{BIAS\_COL}. Pixels read off $60 \times N$ (60, 120, 180 \ldots) pixels 
before a hot pixel are problematic and are also marked as \texttt{BIAS\_COL}. 

The BPM images include static masks that capture persistent systematic issues 
with DECam. Pixels 
within the 15 rows or columns nearest the edge are unsuitable for quantitative 
measurement and marked as \texttt{EDGE}. We have observed several regions in
the flats with persistent structure in the readout direction, including persistent 
defects and smearing. Rectangles around these regions have been manually marked. In addition, 
each hot pixel has a small region around it where the brighter-fatter correction 
will produce anomalous results. Neither of these effects is modeled quantitatively, 
although both appear to be quite small. Pixels affected by these effects
are marked as \texttt{SUSPECT} for exclusion from particularly
precise tasks. Similarly, some columns have significantly
different sky levels than their neighbors but no obvious difference in the bias 
and flat images and are manually marked as
\texttt{FUNKY\_COL}. Regions with an inconsistent response
detected in stacks of sky exposures are marked as
\texttt{WACKY\_PIX}. We do not use amplifier A of DECam CCD 31,
because its gain is unstable at the 10\% level, and it is marked
\texttt{BADAMP}. The star flat correction is unreliable on the 1-2\% level  
within 25 columns of either CCD edge and the data here is marked
\texttt{NEAREDGE}. Finally, each CCD contains six roughly 1.5
mm ``tape bumps'' where two-sided tape was placed to maintain the
relative position of and spacing between the CCD and its aluminum nitride 
readout board \citep{2006SPIE.6276E..08D}. These tape bumps have unstable 
electrostatic properties (compared to other locations on each sensor) that affect
both photometry and astrometry in ways that change with each thermal cycle of
the camera. These regions are marked \texttt{TAPEBUMP} and, similar to \texttt{SUSPECT} regions, 
may not be usable for some applications. Overall, typically 
3\% of pixels are flagged as being on or near the edge of the CCD, and an additional 0.2\% of 
pixels have some other flag that makes them unusable (excluding the unusable half of CCD 31). 
Roughly 1\% of pixels are marked \texttt{TAPEBUMP} or \texttt{SUSPECT}.

We need a sky background model to subtract from our images (Section \ref{sect:skysub}). We use 
a set of 1000 exposures from a given epoch to create an outlier-subtracted sky model defined in terms of four 
Principal Component Analysis \citep[PCA;][]{1986pca..book.....J} components. The first component 
is a median image that includes the pupil effect (a scattered light torus) on large
scales and fringing which is strongest in the $z$ and $Y$ bands on smaller scales. These two effects were handled
with a separate routine in earlier instances of the processing pipeline \citep{Y1A1}. The
second and third components are smooth, orthogonal gradients. The fourth component contains
the pupil and fringe patterns in a different ratio from the first component and allows us to account for variations
in the ratio of airglow to moonlight. Higher order components do not appear to be physically meaningful. 
The PCA components are represented both as full-resolution images over the exposure plane and as lower 
resolution maps consisting of the median of each $128 \times 128$ block of pixels (called a ``superpixel''). Individual 
exposures can be robustly fit with the low resolution images, and the corresponding full resolution background can 
be subtracted off. \citet{Bernstein2017} describes the background model in more detail.

We derive a ``star flat'' that quantifies the differences between the dome flat and the true response to astronomical flux.
These are composed of 22 exposures of 30 seconds each taken in different moderate density star fields. We dither these exposures by 
angles ranging between 10\arcsec\ and 1 degree so that the same star will fall on many different locations 
across the focal plane. We then process these images identically to science exposures up to
this stage and make stellar catalogs using \textsc{SExtractor} \citep{SEx}. We can then use these stars to model the 
spatial and color variation of response of the camera and optical system. \citet{Bernstein2017} describes these star flats in more detail.



%% file: preprocessing/biascor.tex
\subsection{Flat Field Correction and Initial Masking}\label{sect:flatfield}

Having produced a series of image calibration files (see Section \ref{sect:calibration}), we can perform flat field correction and initial pixel masking. 
We use the DES-specific tool, \textsc{Pixcorrect}\footnote{\url{https://github.com/DarkEnergySurvey/pixcorrect}}, to apply our image calibration operations to the crosstalk-corrected images in the following order:
\begin{enumerate}
\item Subtract a bias image
\item Apply a nonlinearity correction
\item Mask saturated pixels
\item Apply a bit mask to the science image
\item Correct bit mask columns where appropriate
\item Apply a gain corrections
\item Apply a brighter fatter correction
\item Divide by the dome flat
\end{enumerate} 

These steps are applied separately to each CCD image of each DECam exposure.
Each CCD image consists of a three-layer FITS file including flux, mask, and weight HDUs. 
The mask layer is initialized to the BPM for that CCD. Later processing steps set bits in the
mask layer for bright star bleed trails, edge bleeds (see below), detected cosmic rays, and detected satellite streaks (the
flux layer is modified in special cases where known bad pixels can be fixed, and to correct effects such as non-linearity,
crosstalk and brighter-fatter). The weight plane layer is initially set to the inverse variance weight calculated from the sum of the variance in the number of 
photoelectrons in each pixel and the variance of the readout noise.\footnote{The number of photoelectrons per pixel is calculated from the ADU per pixel by dividing by the gain of each CCD amplifier} 
This weight is used in initial masking and to fit the sky background. 
After fitting the sky background, the weight layer of each image is replaced by the best-fit sky model in photoelectrons (Section \ref{sect:skysub}). 

Our bias subtraction is a simple pixel-by pixel subtraction of the (crosstalk-corrected) bias image. 
We perform the nonlinearity correction for each pixel using the precomputed CCD-dependent 
lookup table described in Section \ref{sect:calibration}.
We mask those pixels whose values exceed their saturation values with
the \texttt{SATURATE} bit in our bitmask (described below). Saturation
values are different for each amplifier and are noted in the image
headers. Typical values are $\sim$175,000 photo-electrons ($\sim$44,000 ADU). 

We use the BPM images from Section \ref{sect:calibration} to produce an initial bitmask 
for the image. All of our bitmask codes are defined in Appendix \ref{sect:flags}, Table \ref{tab:bitmask}
(prefaced with \texttt{BADPIX}), but we only set 
some of the bits at this processing stage. The remaining mask bits are set by the saturation and 
bleed masking routine (Section \ref{sect:bleedmask}) and the cosmic ray and streak 
masking routine (Section \ref{sect:immask}). 

We initially mark all image bits with a corresponding BPM bit of \texttt{FLAT\_MIN}, 
\texttt{FLAT\_MAX}, \texttt{BIAS\_HOT}, \texttt{BIAS\_WARM}, \texttt{BIAS\_COL}, 
\texttt{FUNKY\_COL} or \texttt{WACKY\_PIX} as \texttt{BPM}. 
Columns marked with BPM bit of \texttt{FUNKY\_COL} and partial columns marked as 
\texttt{BIAS\_COL} (those pixels read out before a hot pixel in their column)
 can often be corrected by subtracting a constant offset as follows. If half of the nearest 
twelve columns (six in either direction) to a \texttt{FUNKY} or partial 
\texttt{BIAS} column are normal, we calculate an outlier clipped mean and variance 
(with an additional linear slope subtracted from the column) for each column. We 
take the average of these values across all comparison columns to produce a ``comparison mean'' 
and a ``comparison variance.'' If the \texttt{BIAS} or \texttt{FUNKY} column's variance 
is within 25\% of the comparison variance, we subtract off the difference between 
the column's mean and the comparison mean so that the column has the same mean value 
as its comparison mean. The \texttt{FUNKY} or partial \texttt{BIAS} is then relabeled 
as \texttt{FIXED} in the image mask. The corresponding \texttt{BPM} bits are then unset. 
Columns without six acceptable neighboring 
columns (among the twelve columns described above) or columns whose variance is more than 25\% 
greater than their comparison variance remain marked as \texttt{BPM}.

The BPM bits of \texttt{EDGE}, \texttt{SUSPECT}, \texttt{NEAREDGE} and 
\texttt{TAPEBUMP} are passed directly into the image mask with the corresponding 
bitmask bit. The actual values of these
flags are different in the BPM and image mask (see Tables \ref{tab:bpm} and 
\ref{tab:bitmask}).

Our final three steps are simple operations. We use our (amplifier-specific) gain correction 
to convert from ADU to electrons. This correction is roughly a factor of 4 for DECam. We apply the brighter-fatter correction (Section 
\ref{sect:calibration}). We finally apply the flat field correction, dividing the image by the 
dome flat values. Mathematically, the calibrated pixel value, CAL, is:
\begin{equation}
\rm{CAL} = \frac{\rm{BF}\left(\rm{Lin}(\rm{RAW}-\rm{BIAS}) \times \rm{GAIN}\right)}{\rm{FLAT}},
\end{equation} 
where Lin is our linearization correction, GAIN is a multiplicative gain constant, BF is our brighter-fatter correction and 
RAW, BIAS and FLAT are (crosstalk-corrected) raw, bias and dome flats, respectively. GAIN is an amplifier-specific conversion 
between the ADU and the more physical unit of photo-electrons.

At this point, we have also made an inverse variance weight image to go along 
with the science image. This weight image includes noise from the bias and flat 
images added in weighted quadrature to the initial image Poisson noise. 
Pixels marked with \texttt{EDGE} and \texttt{BADAMP} are given zero weight value. 

For the first two years of the survey, images were processed slightly differently. 
Particularly, saturation masking was performed
as part of the crosstalk correction, and the brighter-fatter correction was not performed at all. 
All processed data in the public data release, including the reprocessed Y1 and Y2 data, were processed 
with the above methods.

%% file: firstcut/firstcut.tex
\section{First Cut and Final Cut}\label{sect:firstcut}

We use the First Cut pipeline to process all DES exposures within a few hours 
of acquisition so
that they can be evaluated for depth and image quality. Exposures that do
not meet our survey quality criteria are flagged as such, and that area is reobserved 
on a subsequent night. Rather than settling on a static evaluation
pipeline, we have elected to periodically update the First Cut pipeline 
to incorporate improvements as the survey progresses. This strategy was 
crucial for the evolution of the supernova difference imaging pipeline and has 
allowed the DES Collaboration to perform initial investigations using the best 
reduction of the survey data in real time. In practice, the First Cut pipeline 
in use at any time during the survey is similar to the Final Cut (Section \ref{sect:finalcut}) 
used in the latest release production campaign. One crucial difference is that 
First Cut processing uses preliminary image calibrations (Section \ref{sect:preprocess}), 
necessarily composed of data taken before the science exposure,
while Final Cut uses final image calibrations that make use of the entire season. 

The basic outline of the First Cut pipeline is:
\begin{enumerate}
\item Calculate initial astrometry solution (Section \ref{sect:scamp})
\item Mask saturated pixels and associated bleed trails (Section \ref{sect:bleedmask})
\item Fit and subtract sky background (Section \ref{sect:skysub})
\item Divide out star flat (Section \ref{sect:skysub})
\item Mask cosmic rays and satellites (Section \ref{sect:immask})
\item Model the point spread function (Section \ref{sect:psfex})
\item Produce single epoch catalogs (Section \ref{sect:fccat})
\item Evaluate image data quality (Section \ref{sect:eval})
\end{enumerate}
We describe these steps in detail and point out major changes between the
initial First Cut used during SV and Y1 and more modern versions in
the remainder of this section. 
Section \ref{sect:finalcut} describes the small differences between our First Cut and Final Cut processing. The latter is used to produce consistently processed data for our cosmological analyses and public data releases.

%% file: firstcut/astrometry.tex
\subsection{Initial Astrometry}\label{sect:scamp}

We compute a single-exposure astrometric solution using \textsc{SCAMP} 
\citep{Bertin2006} which is part of the \textsc{Astromatic} suite of astronomy 
software.\footnote{\url{https://www.astromatic.net/}}
We generate \textsc{SExtractor}\citep{SEx} catalogs of bright objects that are read by \textsc{SCAMP} and matched the bright sources  in archival reference catalogs with reliable astrometry.
 We use this initial astrometric solution to match our stars to reference catalogs
for initial photometric calibration and PSF modeling. We later improve this solution with multiple 
DES exposures for our final analysis (Section \ref{sect:coadd-astro}). 

The astrometry input catalogs were generated by \textsc{SExtractor} using the settings described in 
Appendix \ref{sect:software1}, Table \ref{tab:scampsex}. These catalogs were generated with a higher detection threshold and less deblending 
than typical science catalogs (e.g., in Section \ref{sect:fccat}) to minimize spurious detections. It also assumes an initial 
seeing FWHM of $1\farcs2$ since the PSF has not yet been modeled. \textsc{SExtractor} requires contiguous pixels 
to be above a detection threshold, and its performance is generally improved if one convolves the image with a small 
kernel to remove noise that dominates at scales smaller than the PSF. To avoid compromising deblending, we convolve our image with a standard $3 \times 3$ 
pixel ``all-ground'' filter (values 4 in the middle, 1 in the corners and 2 on the sides). We use the same filter for all other 
\textsc{SExtractor} uses except for difference image object detection in Section \ref{sect:sndetect} and the final 
coadd catalog in Section \ref{sect:coadd-cat}. This convolution is internal to \textsc{SExtractor}, and the image is 
unchanged for later analysis.

We match these sources across the entire exposure with \textsc{SCAMP} to a reference catalog using the settings  
in Appendix \ref{sect:software1}, Table \ref{tab:scamp}. \textsc{SCAMP} calculates an astrometric 
solution with a TPV World Coordinate System \citep[WCS;][]{WCS}. It takes as input a fixed 
initial-guess set (calculated from a set of earlier exposures) of PV third order WCS distortion terms for each of the 62 CCDs in the DECam array. 
\textsc{SCAMP} also calculates 
the typical astrometric error between our stars and those in the archival reference catalog. In our initial
implementation we used the UCAC4 stellar catalog \citep{2013AJ....145...44Z} as our reference catalog. 
In Y4,  we found that using the 2MASS point source catalog \citep{2006AJ....131.1163S}, 
which has higher source density than UCAC, prevents rare \textsc{SCAMP} failures, and 
we began using it instead. 
We have typical DES-2MASS difference of 0\farcs25. These errors are dominated by 2MASS 
which has typical single detection uncertainties of 0\farcs2 for sources fainter than $K_S$ of 14 which 
we match to. In addition, the proper motion of stars
between the 2MASS epoch and DES epoch is not corrected for and contribute to these differences. Section \ref{sect:coadd-astro} shows 
how we ultimately achieve a relative, internal DES uncertainty of $\approx$ 0\farcs03 by combing overlapping 
exposures. Future processing of DES data with the Gaia catalog \citep{2016A&A...595A...2G} as a reference will allow DES to 
reduce relative and absolute astrometry uncertainties to below 0\farcs03.

%% file: firstcut/initialmask.tex
\subsection{Saturation and Bleed Trail Masking}\label{sect:bleedmask}

In addition to the masking of pixel artifacts described in Section \ref{sect:flatfield}, we also mask saturated pixels, streaks and cosmic rays. 
Here we discuss the masking of bright stars, their associated charge overflow in the readout direction (called ``bleed trails''), and the build up of charge they cause 
near the edge of CCDs. 
Pixels with any of these effects are generally not usable.  We 
describe our procedure for masking satellite streaks and cosmic rays in Section 
\ref{sect:immask}. The flags mentioned here are listed in Appendix \ref{sect:flags}, Table 
\ref{tab:bitmask}.  

Many image-specific artifacts are from bright stars and their associated 
bleed trails. We initially identify bleed trails as saturated columns along the readout 
axis of the image. They are  
caused by charge spreading along the readout columns from a saturated object. Our 
algorithm searches for trails with a $20 \times 1$ pixel kernel (long in the 
readout direction) and requires 7 contiguous saturated pixels to trigger. It then finds all 
saturated pixels adjacent to this kernel to identify an initial trail. This trail is 
typically hundreds of pixels long and includes the bright star. We extend 
this region by 6 pixels orthogonally to the readout direction to compute a rough estimate of the 
area that will be masked. We then produce a rectangle 5 times bigger than this region in 
both the $x$ and $y$ direction. We fit the background in non saturated area of this rectangular region, iteratively 
rejecting $5\sigma$ outliers (using up to 100 iterations). All pixels in a contiguous region 
from the saturated trail down to where the flux goes below $1\sigma$ above the fitted background are 
masked as \texttt{TRAIL}. 
 
We have not yet differentiated the actual saturated star from its corresponding bleed 
trail. Using a stellar center defined by the median bin in our initial saturated trail, 
we fit a circle with a radius such that its outermost pixels have a value that is $3\sigma$ 
above the background (as calculated in the previous paragraph). We mask all pixels within 
2 times this radius as \texttt{STAR}.
If the bleed trail reaches the read registers, we search for deviations in the background 
light near the CCD edge of the amplifier containing the bleed trail. The pixels in a rectangular 
region where the background deviates by more than $1\sigma$ are flagged as \texttt{EDGEBLEED} (see Fig. 5 of \citealt{Bernstein2017}).

Finally, we mask the pixels at the analogous position of a bleed trail in the neighboring amplifier on the same CCD 
as \texttt{SSXTALK}, because the crosstalk signal from saturated pixels is not properly corrected by our linear crosstalk algorithm (Section \ref{sect:crosstalk}).

%% file: firstcut/skysub.tex
\subsection{Sky Subtraction}\label{sect:skysub}

To isolate astronomical sources and remaining artifacts from the field, we 
must first subtract a background component from the image. To do this, we fit and subtract 
off a sky background image using the PCA templates from Section \ref{sect:calibration} 
and divide our image by a nightly star flat.

We describe the production of sky template images in Section \ref{sect:calibration}. We bin our science images
into the same low resolution superpixels as the templates (the median of $128 \times 128$ pixel square). We then fit this 
low resolution image with the 4 corresponding low resolution PCA templates. We multiply our 4 full resolution PCA sky template 
images by these 4 PCA coefficients to produce a full-resolution sky background model. We subtract this image from the science 
image. At this stage, we produce an updated weight image from the inverse variance of the number of photo-electrons in the sky model. 
This avoids a statistical bias in weight in which pixels with positive fluctuations are are given more weight than those 
with negative fluctuations. 

This algorithm for fitting and removing sky for Y3A2 works on the whole DECam exposure at once, rather than
normalizing on a CCD-by-CCD basis as was done in Y1A1. This results in important improvements to the overall
sky fitting and background removal. This is particularly important for detecting extended objects that cover areas more than 1 arcminute
in extent (i.e. extended light from galaxy clusters or very faint extended low surface brightness objects).
We refer the reader to \citet{Bernstein2017} for a full description of the sky subtraction algorithm.

\begin{figure}
  \includegraphics[clip,width=\columnwidth]{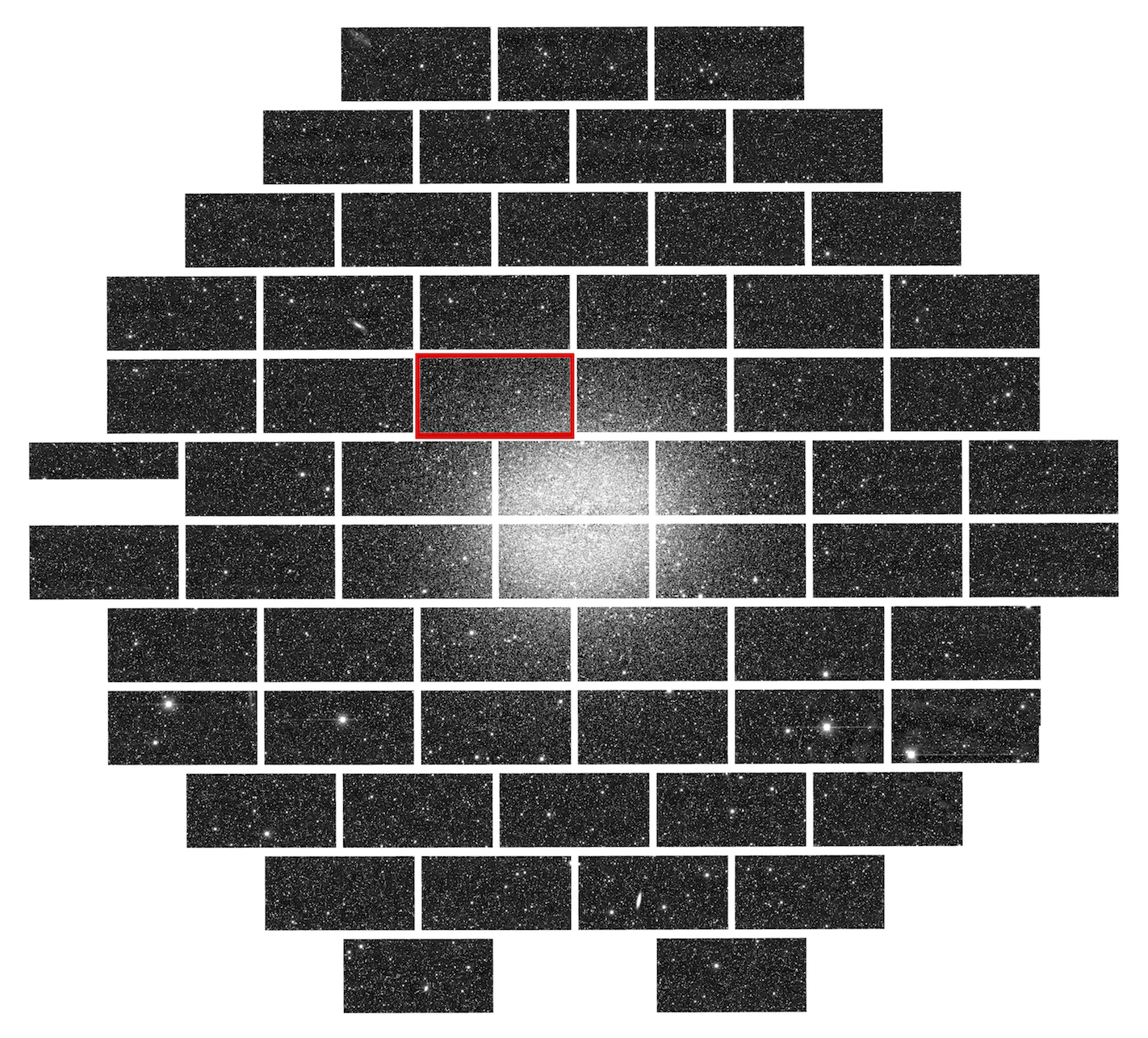}
  \includegraphics[clip,width=\columnwidth]{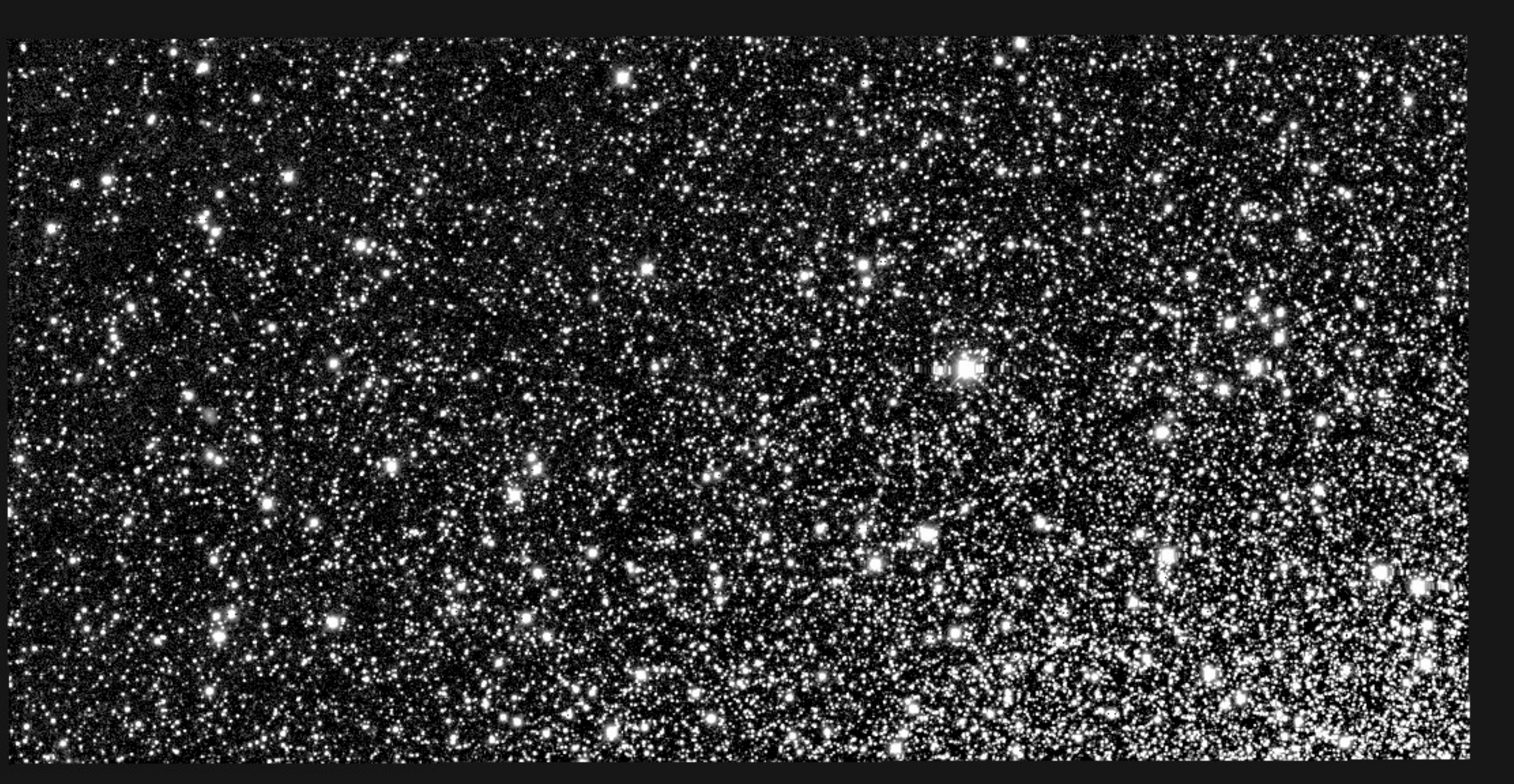}

\caption{DECam exposure number 164189, a 100 second $r$-band image of Omega Centauri. The top panel 
shows the full camera image. CCD 2 is functional in this image, and contains data. CCD 61 and the lower half
of CCD 31 are unusable and blacked out. The bottom panel shows a zoom in of CCD 22, which is boxed in red in the top panel.
}\label{fig:FIRSTCUT}
\end{figure}

At this point, the image is divided by the star flat (Section \ref{sect:calibration}) to correct the differences between 
the dome flat and the true response to stellar flux.
Following division by the star flat, First Cut and Final Cut image
processing is nearly complete. Fig. \ref{fig:FIRSTCUT} shows a sample exposure with all the above processing steps
applied. This is DECam exposure number 164189, a 100 second $r$-band image of the Omega Centauri globular cluster
field.

%% file: firstcut/additionalmask.tex
\subsection{Cosmic Ray and Satellite Masking}\label{sect:immask}

Having subtracted the background image and performed the masking associated 
with the bad pixels in Section \ref{sect:flatfield} and saturated pixels in 
Section \ref{sect:bleedmask}, we now mask our final artifacts: cosmic rays and streaks. 
Masking cosmic rays and streaks in DES was investigated in \citet{2016A&C....16...67D}; however, those algorithms are not applied in First Cut or Final Cut processing.

When cosmic rays interact with the silicon of the DECam CCDs they produce image artifacts that can range 
in appearance from small dots to long streaks. These defects are distinct because they affect a few adjacent 
pixels and are not convolved with the PSF. To detect cosmic rays, we use a variant of the cosmic ray detection algorithm from the LSST code base \citep{2015arXiv151207914J}.
This algorithm examines all pairs of opposing 
pixels (i.e. the north-south pair, the east-west pair, the NW-SE pair and the NE-SW pair) 
surrounding each pixel. We define a cosmic-ray pixel to be $6 \sigma$ above the mean of each of the four pairs (where $\sigma$ is calculated from the local sky background). We also require:
\begin{equation}
I_{pix} - 1.5 \sigma_{pix} > 0.6 P_{pair} (I_{pair} + 1.5 \sigma_{pair}) 
\end{equation} 
for at least one pixel pair. Here, $I_{pix}$ and $\sigma_{pix}$ are the intensity and 
background uncertainty of the pixel in question while  $I_{pair}$ and $\sigma_{pair}$ are 
the mean intensity and uncertainty of the mean for one of the four pairs. $P_{pair}$ is the 
expected average value of the PSF for each pair (with the PSF normalized to unity at the 
central pixel). This second requirement indicates that background gradient is not changing faster 
than the PSF for at least one neighbor pair and takes advantage of the fact that at least 
one neighbor pair will generally be unaffected by the cosmic ray. Pixels that satisfy this requirement as well as their 
neighbors (including diagonal neighbors) are masked with the \texttt{BADPIX\_CRAY} bit.

Streaks are caused by satellites crossing the field of view or linear image defects, and they 
appear as elongated luminous regions often several arcseconds wide and many arcminutes 
long. The Hough Transform \citep{HOUGH} extracts features (typically lines) from an image. 
To perform our Hough transform, we ignore all 
previously masked pixels and dilate our masks around bleed trail masks (Section \ref{sect:bleedmask}). This 
includes diagonal dilation. We then bin the image into $8 \times 8$ superpixels, 
mark superpixels more than $1.5 \sigma$ above the mode and perform a Hough transform on the 
binary mask image of marked and unmarked pixels. We detect peaks in the Hough transform image as 
$18 \sigma$ overdensities. The contiguous area around the Hough peak that exceeds 
$12 \sigma$ is marked as a Hough peak. 
The Cartesian coordinates of this point in Hough space are related to the angle, $\theta$, that a line makes with the image and the line's distance from the center of the image. 
Peaks that cover a $\theta$ range of more than 
$75^\circ$ in Hough space or peaks that are more than 300 pixels wide are ignored to avoid excessive masking. Finally, 
we ignore Hough peaks that have less than 33\% of their pixels exceeding the $1.5 \sigma$ threshold. Real pixels corresponding 
to the remaining Hough peaks are masked with the \texttt{BADPIX\_STREAK}
bit. 

At this stage, we also assign 0 weight to pixels with the \texttt{BADPIX} mask bit \texttt{BADAMP}, 
\texttt{EDGEBLEED} or \texttt{EDGE} or \texttt{CRAY} set. Before Y3 processing, masked cosmic ray 
regions were interpolated across during this masking. But after Y3 
(including all Y1 and Y2 data reprocessed for public release), this feature was removed as these 
regions have zero weight.


%% file: firstcut/psfex.tex
\subsection{PSF Extraction}\label{sect:psfex}

We model the point spread function (PSF) of each image individually
using \textsc{PSFEx} \citep{PSFEx}, a PSF analysis package part of the
\textsc{Astromatic} suite of astronomy software. The first step in
using \textsc{PSFEx} is to make a ``bright star'' catalog for each CCD
image using \textsc{SExtractor} and the parameters in Appendix \ref{sect:software1}, 
Table \ref{tab:psfexcat}. Parameter values differ slightly from those used in Y1. 
We convolve the image with the same $3 \times 3$ pixel ``all-ground'' convolution mask 
from Section \ref{sect:scamp}.

We use these bright star catalogs and their accompanying images as
inputs for \textsc{PSFEx} using the parameters in Appendix \ref{sect:software1}, Table
\ref{tab:psfex}.  \textsc{PSFEx} takes unsaturated point sources 
from the catalogs, examines each object and iteratively models the PSF 
while excluding non-PSF sources. Ultimately, \textsc{PSFEx} produces a model of the 
PSF across each CCD of every on-sky exposure.  For Y3, the model varies with degree 2 
(quadratically) across each $4096\times2048$ CCD. The PSF modeling described here is 
used for the photometry in our official data release. Some DES analyses (e.g. weak 
lensing described in \citealt{2017arXiv170801533Z}) use different PSF modeling. 

%% file: firstcut/catalog.tex
\subsection{Single Epoch Catalog Production}\label{sect:fccat}

We use \textsc{SExtractor} to produce source catalogs for individual images. The 
parameters we use are described in Appendix \ref{sect:software1}, Table \ref{tab:firstcutsex} 
and we convolve the data with same $3 \times 3$ pixel ``all-ground'' filter from Section 
\ref{sect:scamp}. We use the fitted PSF models described in Section \ref{sect:psfex}.
The measured parameters include those described in Table \ref{tab:cat}.
The detection threshold for the Final Cut catalogs corresponds to ${\rm S/N \sim 10}$. 
We note that the output catalogs include the parameter \texttt{IMAFLAGS\_ISO}, which expresses the union of all mask bits present for pixels within the isophotal radius of each source.

\begin{table}
\centering
\begin{tabular}{ll}
        \hline
Parameter & Description \\
        \hline
\texttt{ALPHAWIN\_J2000} & Right ascension\\
\texttt{DELTAWIN\_J2000} & Declination\\
\texttt{FLAGS} & Detections flags\\
\texttt{FLUX\_AUTO} & Kron flux\\
\texttt{FLUXERR\_AUTO} & Kron flux uncertainty\\
\texttt{FLUX\_PSF} & PSF flux\\
\texttt{FLUXERR\_PSF} & PSF flux uncertainty\\
\texttt{A\_IMAGE} & Major elliptical axis\\
\texttt{B\_IMAGE} & Minor elliptical axis\\
\texttt{THETA\_IMAGE} & Elliptical position angle\\
\texttt{IMAFLAGS\_ISO} & Isophotal image flags\\
\texttt{SPREAD\_MODEL} & Non-PSF flux parameter\\
\texttt{SPREADERR\_MODEL} & Non-PSF flux error\\
        \hline
\end{tabular}
\caption{Output \rm{\textsc{SExtractor} quantities reported in single epoch catalogs.}}\label{tab:cat}
\end{table} 

%% file: firstcut/eval.tex
\subsection{Image Evaluation}\label{sect:eval}

DES used historical observing conditions at CTIO to establish a series of survey requirements 
which translate into requirements on the PSF, background and atmospheric transmission that each 
exposure must satisfy to be included in the final wide-area survey. We combine these three 
quantities into a value called $t_{\rm eff}$, which we use to evaluate image quality \citep{Teff}. 
We also impose separate requirements on the PSF and sky background.

$t_{\rm eff}$ is the ratio between the actual exposure time and the exposure time necessary to achieve the same signal-to-noise for point sources observed in nominal conditions.
An exposure taken under ``fiducial'' conditions has $t_{\rm eff} \equiv 1$. These fiducial conditions (Table \ref{tab:eval}), are above average, but not uncommon for CTIO.
 Specifically, we define
\begin{equation}
t_{\rm eff} \equiv \left(\frac{FWHM_{fid}}{FHWM}\right)^2\left(\frac{B_{fid}}{B}\right) F_{trans}.
\end{equation}
$FWHM$ and $FWHM_{fid}$ are the measured and fiducial PSF full width half max, respectively. 
$B$ and $B_{fid}$ are the measured and fiducial sky background. $F_{trans}$ is the atmospheric 
transmission relative to a nearly clear night.

\begin{table}
\centering
\begin{tabular}{ccccc}
        \hline
       & $FWHM_{fid}$ & $B_{fid}$         & Minimum   & $FWHM_{max}$ \\
Filter & (arcseconds) & (electron s$^{-1}$) & $t_{\rm eff}$ & (arcseconds)\\
        \hline
$g$    & 0.9927    & 1.05  & 0.2 & 1.765\\
$r$    & 0.9369    & 2.66  & 0.3 & 1.666\\
$i$    & 0.9       & 7.87  & 0.3 & 1.6\\
$z$    & 0.8685    & 16.51 & 0.3 & 1.544\\
$Y$    & 0.8550    & 14.56 & 0.2 & 1.520\\
        \hline
\end{tabular}
\caption{The fiducial and limiting values for parameters used in our First Cut image evaluation. 
}\label{tab:eval}
\end{table}

$FWHM_{fid}$ are listed in Table \ref{tab:eval}. To obtain these values, we multiply a $FWHM$ of 0\farcs9 in the $i$ band by a coefficient calculated with a Kolmogorov turbulence model (\citet{1981PrOpt..19..281R}, 
\citet{1941DoSSR..30..301K}) that converts between $i$-band seeing and the seeing that would be observed in 
another band. We have calculated a set of DECam sky brightnesses from a ``photometric'' night (2012 December 15)
to obtain our fiducial $B$, listed in Table \ref{tab:eval}. 

We calculate atmospheric transmission as a robust, outlier-rejected median difference between the 
First Cut stellar magnitudes and those from well-calibrated datasets. We compare DES $i$, $z$ and $Y$ 
magnitudes to 2MASS point source $J$ magnitudes \citep{2006AJ....131.1163S} accessed via the Naval Observatory 
Merged Astrometric Dataset \citep[NOMAD;][]{2004AAS...205.4815Z}. We compare the DES $g$- and $r$-band magnitudes to their 
counterparts in the The AAVSO Photometric All-Sky Survey \citep[APASS;][]{2009AAS...21440702H}  when these 
catalogs are deep enough and a combination of NOMAD $J$ and $B$ band \citep{2003AJ....125..984M} otherwise. 
We convert this magnitude difference, $\Delta m$, to a multiplicative transmission factor:
\begin{eqnarray}
F_{trans} &=& 1,\ \rm{for}\ \Delta m \leq 0.2, \nonumber\\
F_{trans} &=& 10^{-0.8 (\Delta m-0.2)},\ \rm{for}\ \Delta m > 0.2.
\end{eqnarray}

We reject exposures with $t_{\rm eff}$ less than the minimum values listed in Table \ref{tab:eval}. 
In practice, this is usually the most stringent image requirement. We also require $i$-band seeing to be less 
than $FWHM_{max}$ (Table \ref{tab:eval}). $FWHM_{max}$ was constructed using the same Kolmogorov 
conversions as $FWHM_{fid}$ and the constraint that the $i$-band PSF have $FWHM <$ 1\farcs6.  

\textsc{ObsTac} uses onsite monitors to estimate sky brightness and seeing before an image is taken. 
If \textsc{ObsTac} estimates that the $i$-band seeing at zenith does not satisfy $FWHM_i <$ 1\farcs1, it schedules observations of supernova fields that have not been observed in four or more nights. 
If \textsc{ObsTac} estimates that the sky background may exceed acceptable limits, it does not schedule observations in $g$ or $r$ band, and, in extreme cases, it restricts observations to the $Y$ band.

%% file: firstcut/finalcut.tex
\subsection{Final Cut}\label{sect:finalcut}

As mentioned above, we process each wide-area exposure at least 
twice: First Cut processing is run within a few hours of acquisition and
Final Cut processing is run months later to assemble a data release with an improved version of the software.

In addition to the relatively minor code changes between the First Cut and the Final Cut pipelines, 
we also used different calibration images. Specifically, during the first three years of First Cut 
processing, we calibrated the images using nightly calibration images. All Final Cut processing as 
well as our First Cut processing after Y3 use super-calibration images (Section \ref{sect:calibration}) 
with larger number of input images for more statistical power. The processing of calibration images themselves is also updated 
analogously to the processing of their corresponding science exposures. 

Final Cut data is photometrically calibrated using Forward Global Calibration Method \citep[FGCM;][]{FGCM}. 
This method uses overlapping exposures and a large network of calibration stars 
to calibrate the entire survey while also modeling the time-dependent atmospheric and instrumental 
passbands for each CCD. FGCM assigns each exposure a zero point with typical uncertainty of 7 mmag.

There will be at least three DES data releases internal to the DES Collaboration
(Y1, Y3 and eventually Y5). Each release includes all the data up to that point. The 
DR1 public release includes all data through Y3, and the DR2 public release will include all data up to Y5 (all DES data). For each release, every exposure taken at that time  
is processed with the current Final Cut pipeline. There were significant software 
changes between the beginning of the survey (SV, Y1) and the Y3 data release, the 
first data release with complete coverage of the survey footprint. We anticipate 
additional incremental improvements through Y5. As the processing software
asymptotically approaches its final form, the First Cut and Final Cut
pipelines merge, with the Y5 First Cut being essentially identical to
Y4 Final Cut.

%% file: supernova/supernova.tex
\section{Supernova Processing}\label{sect:supernova}

Detecting and characterizing Type Ia supernovae is one of the primary goals of 
DES. The wide-area cadence includes two observations of each area per band per 
season, which is not adequate for transient light curve characterization. Instead, DES 
repeatedly visits ten 3 deg$^2$ fields when \textsc{ObsTac} estimates that observation 
conditions are marginal (PSF FWHM worse than 1\farcs1 when transformed to $i$-band zenith) or when any supernova field has not 
been observed in seven nights. Roughly 45\% of supernova data is taken in conditions that 
fail the First Cut quality requirements described in Section \ref{sect:eval}.

\begin{figure*}
\centerline{
\includegraphics[width=7.3in]{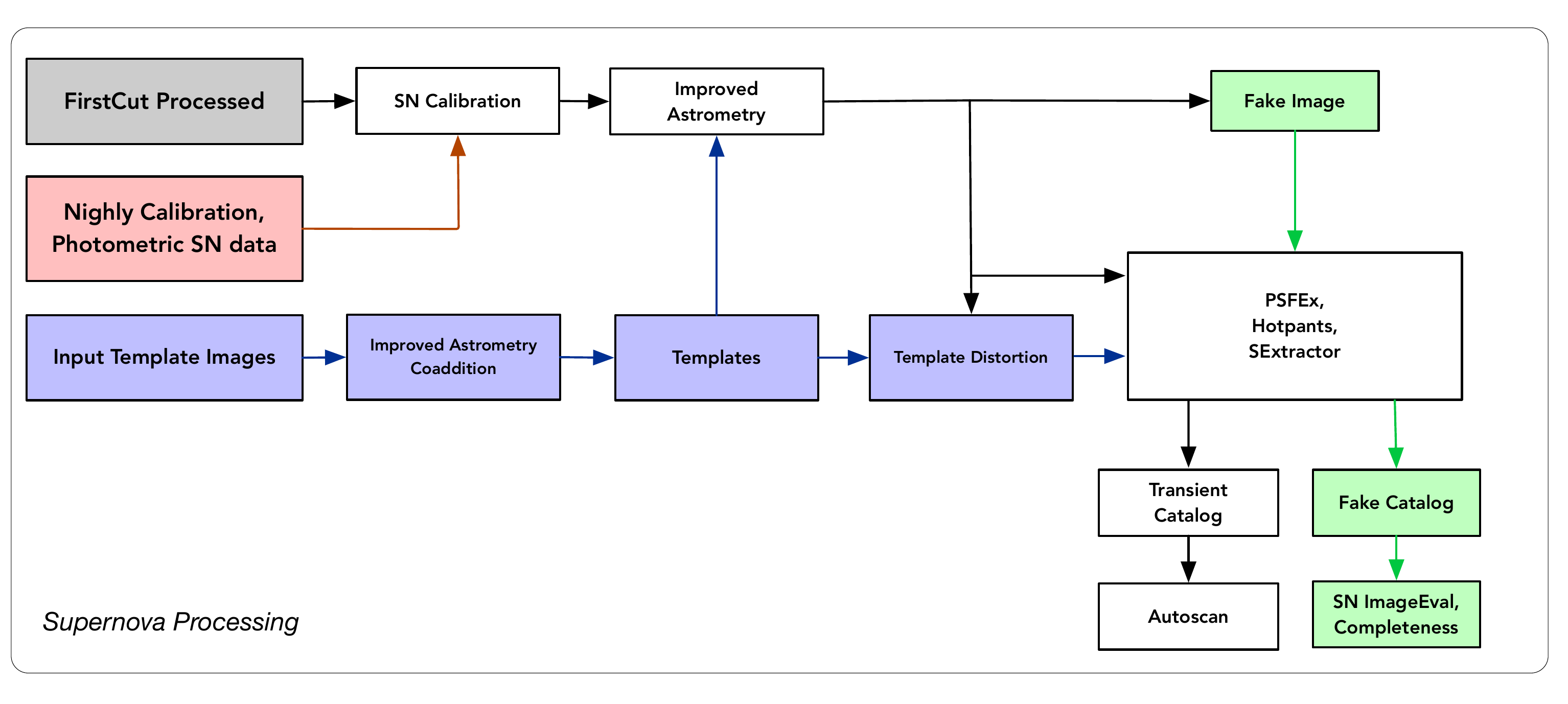}
}
\caption{A diagram of the difference image processing used to detect supernovae and other transient in the DES supernova fields.}\label{fig:supernova}
\end{figure*}

To cover a range of supernova luminosities and implied redshifts, we observe 8 ``shallow'' fields (C1, C2, 
E1, E2, S1, S2, X1 and X2) and 2 ``deep'' fields (C3 and
X3). Table \ref{tab:snfields} shows the center 
of each field. These fields were chosen to avoid bright stars and to overlap with other significant surveys. 
The C fields overlap with the Chandra Deep Field South \citep{2011ApJS..195...10X}. The E fields overlap 
with the ELAIS S1 field \citep{1999MNRAS.305..297G}. The S fields overlap with SDSS stripe 82 
\citep{2007ApJS..172..634A,DR7}. The X fields overlap with the XMM-LSS fields \citep{2004JCAP...09..011P}. 
The exposure times  of each field in each band are in Table \ref{tab:exptime}. The shallow-field $g$, 
$r$ and $i$ observations in a given night are single exposures. The shallow $z$ observation and 
all deep observations are coadditions of exposures taken in uninterrupted sequences. Going forward, we refer to an individual night's 
observation as an ``epoch'' regardless of the number of exposures taken.

\begin{table}
\centering
\begin{tabular}{cccc}
        \hline
      & Deep or &  Field Center (deg)      \\
Field & Shallow & R.A. & Decl.\\
        \hline
C1  & shallow & 54.2743 & -27.1116\\
C2  & shallow & 54.2743 & -29.0884\\
C3  & deep    & 52.6484 & -28.1000\\
E1  & shallow &  7.8744 & -43.0096\\
E2  & shallow &  9.5000 & -43.9980\\
S1  & shallow & 42.8200 &   0.0000\\
S2  & shallow & 41.1944 &  -0.9884\\
X1  & shallow & 34.4757 &  -4.9295\\
X2  & shallow & 35.6645 &  -6.4121\\
X3  & deep    & 36.4500 &  -4.6000\\
        \hline
\end{tabular}
\caption{The location of each supernova field in right ascension and declination.}\label{tab:snfields}
\end{table}

We process the supernova fields with essentially the 
same pipeline as the wide-area First Cut pipeline up through the second masking phase (Section 
\ref{sect:immask}). After an image has been processed, we refine its astrometric solution 
and match it to a template image of the same field. We then subtract the template image from the 
nightly science image, detect sources in the resulting difference image, cross-correlate difference 
detections across different images and evaluate our transient detection efficiency by analyzing 
artificially inserted sources in the images. We diagram this pipeline in Fig. \ref{fig:supernova} and 
summarize it in this section. \citet{2015AJ....150..172K} describes the DES supernova program and 
algorithms in more detail. Here we give an operational description on how our software is run and 
how the supernova processing fits into the context of DES processing.

%% file: supernova/sncal.tex
\subsection{Supernova Calibration}\label{sect:sncal}

Supernova characterization for spectroscopic follow-up requires more precise photometric 
calibration than the typical First Cut wide-area processing. We facilitate the First Cut 
photometric calibration by imaging standard star fields during both evening and morning twilight 
\citep{2007ASPC..364..187T}. These standard star fields are generally in SDSS Stripe 82 or at 
declination $\approx -45^\circ$ \citep[Appendix A of][]{Y1A1}. We use stars from these fields (called secondary standard 
stars) to determine nightly zero points, atmospheric extinction coefficients and color terms. 
We then use a set of ``tertiary standard stars'' within each of the supernova fields for 
photometric calibration. The fields and the tertiary standards within them were observed in photometric 
conditions during the Science Verification period, and their magnitudes and colors are well-measured relative to 
the nightly standards. In good conditions, we typically have 100--200 
tertiary standards per CCD in each field. Relative and absolute photometric calibration have been 
confirmed to the 2\% level in the difference imaging pipeline \citep{2015AJ....150..172K}, and calibration 
continues to improve for the cosmology analysis \citep{FGCM}. 

%% file: supernova/template.tex
\subsection{Supernova Template Images and Improved Astrometry}

Our supernova detection algorithm depends on precise image subtraction between a nightly science image 
and its corresponding template image. This ``difference imaging'' requires template images that are 
astrometrically aligned to the science image.  

We produce a coadded template image per season for each field from exposures from a different season. The Y1 
templates were constructed from science verification data for the initial detection and then remade with Y2 
data for later reprocessing. The Y2 templates were made with Y1 data, and all later templates were made from Y2 data.
For each CCD of each template, we find the image with the lowest sky noise, $\sigma_{\rm{SKY}}^{\rm{min}}$, and then make a 
coadded image from the epochs with the sharpest PSF that have sky noise less than $2.5 \sigma_{\rm{SKY}}^{\rm{min}}$. 
The coadd can include up to 10 epochs, though the average contains 8.4 that satisfy the noise 
requirement. 

We find our template image astrometric solution in two steps. First, each exposure in the template is aligned 
with \textsc{SCAMP} (Section \ref{sect:scamp}) to the the USNO-B1 catalog \citep{2003AJ....125..984M} with 
precision of roughly 100 mas. We calibrate the images (see Section \ref{sect:sncal}) and make a weighted 
intermediate coadd of these exposures. We use the resulting catalogs of stars to make a new reference catalog of relative 
astrometric standards. We re-align the constituent exposures to this coadd catalog, resulting in 20 mas 
(0.08 pixel) relative precision. In the nightly supernova image processing, we use the same coadd reference 
catalog to improve the astrometric solution, typically achieving relative astrometric precision between the 
single epoch image and template of roughly 25 mas (0.1 pixel). In addition to using an improved reference catalog, the supernova 
pipeline (which was developed independently of the First Cut pipeline) uses different astrometry catalog \textsc{SExtractor} 
settings (Appendix \ref{sect:software2}, Table \ref{tab:snscampsex}) and \textsc{SCAMP} settings (Appendix 
\ref{sect:software2}, Table \ref{tab:snscamp}).  

%% file: supernova/psfex.tex
\subsection{Supernova PSF Modeling}\label{sect:snpsfex}

The PSF modeling in the supernova pipeline uses \textsc{PSFEx} but was developed independently of the analogous 
routines in the wide-area pipeline. We use \textsc{SExtractor} and the settings in Appendix \ref{sect:software2}, 
Table \ref{tab:snpsfexcat} to make a catalog of sources for \textsc{PSFEx} analysis. We also 
use the ``all-ground'' filter from Section \ref{sect:psfex}. In addition, we match the sources detected 
here to a previously defined list of pointlike objects constructed from high quality Science Verification data 
in the supernova fields. We then model the PSF with \textsc{PSFEx} and the settings in Appendix 
\ref{sect:software2}, Table \ref{tab:snpsfex}. 

%% file: supernova/detection.tex
\subsection{Detecting Transients in Difference Images}\label{sect:sndetect}

We detect transients as point sources in the difference images. To produce these 
images, we astrometrically distort the template images to match the corresponding single 
epoch image, model and match the PSFs in the science and template images and finally detect sources in their 
difference.

We astrometrically resample the template images using \textsc{SWarp} \citep{Bertin2002}, 
part of the \textsc{Astromatic} software package. Using the astrometry headers created by 
\textsc{SCAMP} (Section \ref{sect:scamp}), and settings in Appendix \ref{sect:software2}, Table \ref{tab:SWarp}, 
\textsc{SWarp} matches the template image to the science image from each epoch on a pixel 
by pixel level. Alignment is typically accurate to roughly 0.1 pixels. At this point, we 
also use \textsc{SWarp} to produce nightly supernova coadd images (shallow field $z$-band images 
and all deep field images). The \textsc{SWarp} settings for coaddition are nearly identical to 
those in Table \ref{tab:SWarp} except that \texttt{RESAMPLING\_TYPE} = \texttt{LANCZOS3} 
and \texttt{PROJECTION\_ERR} = $0.001$.

We photometrically match the (astrometrically-matched) template image to the science image using 
\textsc{HOTPANTS} \citep{2015ascl.soft04004B} and the parameters in Appendix \ref{sect:software2}, 
Table \ref{tab:hotpants}. Briefly, \textsc{HOTPANTS} divides a CCD image into $200$ ($10 \times 20$) 
stamps. Within each stamp it detects stars and estimates the optimal kernel needed to convolve the 
template image of the star into the science image. \textsc{HOTPANTS} then uses these measurements to 
model a spatially varying kernel across the entire image which it in turn uses to produce a 
convolved template image. This kernel includes a spatially-varying scaling term to match the images 
photometrically. The convolved template and the science image are now ideally matched in astrometry, 
photometry and PSF. \textsc{HOTPANTS} outputs a kernel model and difference image which should 
ideally be consistent with zero excluding transients.

We find the point sources for the difference image using \textsc{SExtractor} 
and the parameters in Appendix \ref{sect:software2}, Table \ref{tab:snsex}. \textsc{SExtractor} filters 
(convolves) the difference image using a $5 \times 5$ pixel Gaussian with a FWHM of 3 pixels. The 
resulting catalog of transient detections contains significant contamination and requires automatic 
artifact rejection (Section \ref{sect:caneval}).

%% file: supernova/dofake.tex
\subsection{Fake Images}\label{sect:snfakes}

To assess our transient detection efficiency we add simulated transients, ``fakes,'' to each supernova
exposure. The transients are inserted as point sources (as modeled by \textsc{PSFEx}). There are $\approx 20$ 
fakes per CCD. Only a negligible fraction of galaxies have a fake in any given image, and fakes have a correspondingly 
small impact on transient completeness.

We add two classes of fake sources to the images: isolated 20th magnitude sources and sources 
that simulate a set of Type Ia supernova that span the full range of redshifts and supernova-host separations 
\citep[as per \textsc{SNANA};][]{2009PASP..121.1028K,2015AJ....150..172K}. We use the 20th magnitude  
fakes to evaluate image depth and determine if an image needs to be retaken (Section \ref{sect:sneval}). 
We use the Type Ia supernova fakes to measure detection efficiency versus signal-to-noise (needed for simulations), 
and to monitor the trigger efficiency versus redshift.

%% file: supernova/caneval.tex
\subsection{Transient Candidate Evaluation}\label{sect:caneval}

Difference images are prone to unique artifacts, and we use 
more stringent criteria than in conventional astronomical images 
to ensure that the difference image point sources are real. Difference 
image point sources must satisfy a series of pixel and catalog-based 
requirements. We make postage stamps of sources that pass these cuts and pass these
 on to a more rigorous selection routine called \textsc{autoScan}, described 
below.  

To eliminate the bulk of artifacts, we apply basic quality cuts. Potential 
transient candidates must be brighter than 30th magnitude, have signal to 
noise greater than 3.5 (typically less stringent than the 5$\sigma$ 
\textsc{SExtractor} requirement) and be consistent with a point source PSF (as 
measured by the \textsc{SExtractor} parameter \texttt{SPREAD\_MODEL}). In a 
$35 \times 35$ pixel stamp around the candidate, there must be fewer than 
200 pixels $2\sigma$ below the mean, 20 pixels $4\sigma$ below the mean 
and 2 pixels $6\sigma$ below the mean. This helps reject ``dipoles'' 
(negative regions abutting positive regions), a common form of artifact in 
difference images caused by astrometric misalignment. We also reject objects 
coincident with known photometric variables (including quasars) 
and bright stars from a previously-defined veto catalog. We enter all sources 
that pass these cuts into a database that includes the transient detection from 
previous images. 

Difference sources that pass the above criteria are saved as $51 \times 51$ pixel stamps and 
sent to \textsc{autoScan}, a more rigorous selection algorithm described 
in \citet{2015AJ....150...82G}. \textsc{autoScan} is a machine learning 
selection algorithm that was trained with a set of 900,000 transient detections. 
Half are synthetic transients (the Supernova Ia ``fakes'' described in Section 
\ref{sect:snfakes}) while the other half are human-verified artifacts. 
\textsc{autoScan} uses 37 high level features to distinguish between real detections and 
artifacts. The three most important are the ratio of PSF-fitted flux to 
aperture flux, the magnitude difference between a transient and its 
nearest static neighbor and the aforementioned \textsc{SPREAD\_MODEL}. 
Each source is given a score between 0 (artifact) and 1 (high-quality detection). Two detections 
with a score above 0.3 from different images (from different nights and/or different bands) 
separated by less than 1\arcsec\ form a transient candidate. Detections from additional epochs 
are added to this candidate and analyzed to produce transient light curves and locate 
likely host galaxies \citep{2015AJ....150..172K}, but that process is beyond the scope of this paper.

%% file: supernova/sneval.tex
\subsection{Supernova Image Evaluation}\label{sect:sneval}

To evaluate whether a transient is a Type Ia supernova, we must obtain regularly 
sampled light curves from high quality images. We thus evaluate each supernova image 
within hours of acquisition to determine if it passes our quality cuts. If any of 
the $g$, $r$, $i$ or $z$ images from the shallow fields fail quality cuts, we resubmit the entire 
four image sequence for reobservation. The images from the deep fields are evaluated  
individually (as single-band coadded images), and if any of them fails our quality cut,  
that single band sequence is retaken.

Images are evaluated via two criteria: depth and image quality. We evaluate depth 
by obtaining a mean signal to noise ratio of the magnitude 20 fake sources from 
Section \ref{sect:sneval} in each image (excluding those flagged as being bad 
detection due to chip gaps, etc.). In shallow fields, these sources must have a signal 
to noise ratio greater than 20 while in deep fields they must have signal to noise 
ratio greater than 80. We evaluate the PSF as the mean FWHM of all well-detected fakes. 
We then estimate what the $i$-band PSF at zenith as:
\begin{equation}
\rm{FWHM}_{i, \rm{zenith}} = \frac{\rm{FWHM}}{K \times \rm{AIRMASS}^{0.6}}
\end{equation} 
where $K$ is the appropriate Kolmogorov coefficient (discussed in Section \ref{sect:eval}) 
that converts between the observation band and the $i$ band, and the airmass term 
corrects for the fact that the telescope is not generally pointing at zenith. Supernova 
fields that do not satisfy $\rm{FWHM}_{i, \rm{zenith}} <$ 2\arcsec are sent back to 
the observing queue to be retaken.

%% file: coadd/coadd.tex
\section{Multi-Epoch Processing}\label{sect:coadd}

Coadding exposures to produce deeper images is crucial for most of the DES science 
goals. We do not repeat the masking, calibration or other image processing 
described in Sections \ref{sect:preprocess} and \ref{sect:firstcut}. We first divide 
the sky into ``coadd tiles'' and work with all acceptable exposures within each tile. We optimize 
the coadds by refining the astrometry from Section \ref{sect:scamp}. We then 
resample the individual exposures, assemble the coadded images, 
model the coadded PSF and finally create output catalogs. We produce 
additional files that include pixel-level data and PSF models of every image 
of every detected source and use the perform optimal photometry and morphology 
measurements. We discuss coadd production below. 

%% file: coadd/coadd-tiles.tex
\subsection{Coadd Tiles}\label{sect:coadd-tiles}

Since we cannot process all DES data at once, we divide the sky into coadd tiles that are rectangular tangent projections. Each tile is 
$10,000 \times 10,000$ pixels. Each coadded pixel is a 0\farcs263 square, meaning each tile is 2630\arcsec\ or 
0.7306$^\circ$ on a side. Tiles are arranged along rows of constant declination, with the northernmost 
row being centered at declination 30$^\circ$ and each subsequent row being 0.7138$^\circ$ south. This northern 
declination limit is 25$^\circ$ above any DES data, and it allows us to include non-DES DECam data outside the DES area in our database. Each row has a 
tile centered at right ascension of 275$^\circ$ and tiles are spaced by roughly 0.71$^\circ$ in angular distance in 
either direction. There is a roughly 1\arcmin\ overlap between tiles, so every source will have a tile in which it 
is at least 0.\arcmin5 from the edge. The coaddition processes described below use every image (at the individual CCD level) 
with a given coadd tile.

%% file: coadd/coadd-astro.tex
\subsection{Astrometric Refinement}\label{sect:coadd-astro}

Astrometric solutions of overlapping exposures can vary relative to one another. To 
produce a consistent solution and minimize the width of the coadded PSF, we recalculate our 
astrometric solutions simultaneously. This requires that we find every image (at the CCD 
level) in a coadd tile and rerun \textsc{SCAMP} on the images simultaneously.

For each coadd tile we use the Final Cut \textsc{SExtractor} catalogs for
all of the CCDs in all of the exposures that fall within that
tile for astrometric remapping. The positions of objects in these catalogs already reflect the
Final Cut astrometric solution. We use these catalogs as
input for the multi-epoch astrometric refinement step in which we run
\textsc{SCAMP} (using the 2MASS stellar catalog as our reference for Y3A2 release) and compute a global astrometric solution for all overlapping exposures
using the 
settings in Appendix \ref{sect:software3}, Table \ref{tab:coadd-scamp}.
This results in 0\farcs03  RMS or better relative astrometry. The proper motion of 
Milky Way stars and the final registration of the coadd astrometry to 2MASS, from year 
1999, compared with DES Y1-Y3, obtained during years 2013-2016, cause 
an apparent systematic shift of positions in the DES Y3A2 release 
when compared with Gaia (for example) of up to 350 mas.  Future releases (Y5, DR2) will calibrate on the Gaia 
system to remove 
this offset. Gaia is a much closer temporal match to DES and will enable relative and absolute measurements of star positions of order
0\farcs03 or better.

%% file: coadd/coadd-swarp.tex
\subsection{Coadd Image Generation}\label{sect:coadd-swarp}

Having used \textsc{SCAMP} to recalculate mutually consistent astrometric solutions for the input 
images, we resample the images so that their pixels are matched to those in 
our coadd tile. We perform some preprocessing to ensure only good pixels (as described by their pixel flags) 
are used as inputs into the final coadd. We then use \textsc{SWarp} \citep{Bertin2002} from the 
\textsc{Astromatic} package to resample the input images and produce the final coadd images. 

Inputs to \textsc{SWarp} are the set of all single-epoch Final Cut images which overlap the tile 
in question along with a photometric zeropoint for each CCD for each image.  These photometric zeropoints
come from FGCM described in Section \ref{sect:finalcut}. The set of single epoch input images actually consists of four planes of data. 
The first contains on-sky flux (electrons/pixel). Within the coadd pipeline, we linearly interpolate across columns 
flagged with \texttt{BPM} or \texttt{TRAIL} (but not \texttt{EDGE}) flags to produce contiguous pixels. These interpolated pixels are flagged 
with the \texttt{INTERP} flag. The second plane is a sixteen-bit mask of bad pixels, cosmic rays, 
satellite streaks, bleed trails, etc. The third is a weight plane derived from the fitted sky PCA background for that image.
We use the mask plane from each input image to ensure that dubious pixels around saturated regions 
and other image defects are not included in the final coadd. Specifically, we set the weight of  
pixels with the \texttt{BADAMP}, \texttt{EDGEBLEED}, \texttt{SSXTALK}, \texttt{EDGE} or \texttt{STREAK} flags from Table \ref{tab:bitmask} to zero.
In a fourth plane, we also set the weight of the interpolated \texttt{BPM} and \texttt{TRAIL} pixels to zero. 

Some data are excluded from coaddition either at the exposure or CCD level \citep{Y1A1}. This ``blacklist'' 
of poor data includes CCDs with large scattered light artifacts, spurious excess noise or airplane 
or meteor streaks. It contains fewer than 1\% of CCD images.

We use \textsc{SWarp} and the settings in Appendix \ref{sect:software3}, Table \ref{tab:coadd-swarp} 
to resample and combine the images into the final coadded image. We run \textsc{SWarp} 
once with each of the two weight planes described above. We use the first (less masked) weight plane 
to produce the science image. These images are contiguous across interpolated regions (e.g. bright stars) rather than 
having zero value in these regions. This is both aesthetically pleasing and 
prevents sharp edges around masked regions that produce false detections in \textsc{SExtractor}. 
We use the \textsc{SWarp} weight output, constructed using the second set of weight images, to 
produce the final coadded weight image. This ensures that all coadded pixels have the correct weight, particularly that
pixels with no valid input pixels have zero weight. 

In addition to performing this coaddition on $g$, $r$, $i$, $z$ and $Y$ images separately, 
we also produce a \textsc{CHI-MEAN} detection image which is a \texttt{CHI-MEAN} coadd of the $r$, $i$ and $z$ images \citep{1999AJ....117...68S,Y1A1}. 
We do not include $g$-band images, because they have larger PSFs and 
noisy sky backgrounds that would degrade the detection image. In addition, the DES cosmology 
projects are focused on redder objects. We do not include $Y$-band images, because they have limited depth and irregular PSFs that would also degrade the detections image.  We ultimately run 
\textsc{SExtractor} on this detection image to obtain a master list of sources and source shapes. 

\begin{figure}
  \includegraphics[clip,width=\columnwidth]{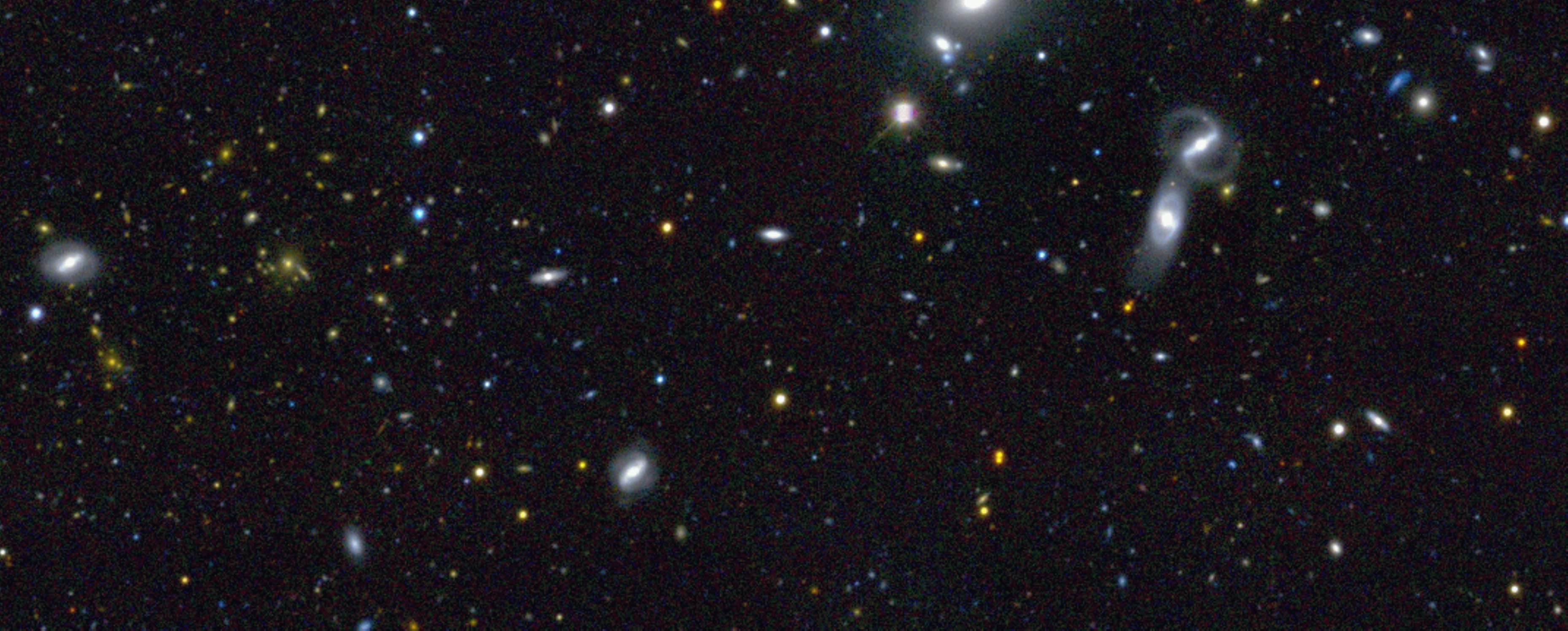}
  \includegraphics[clip,width=\columnwidth]{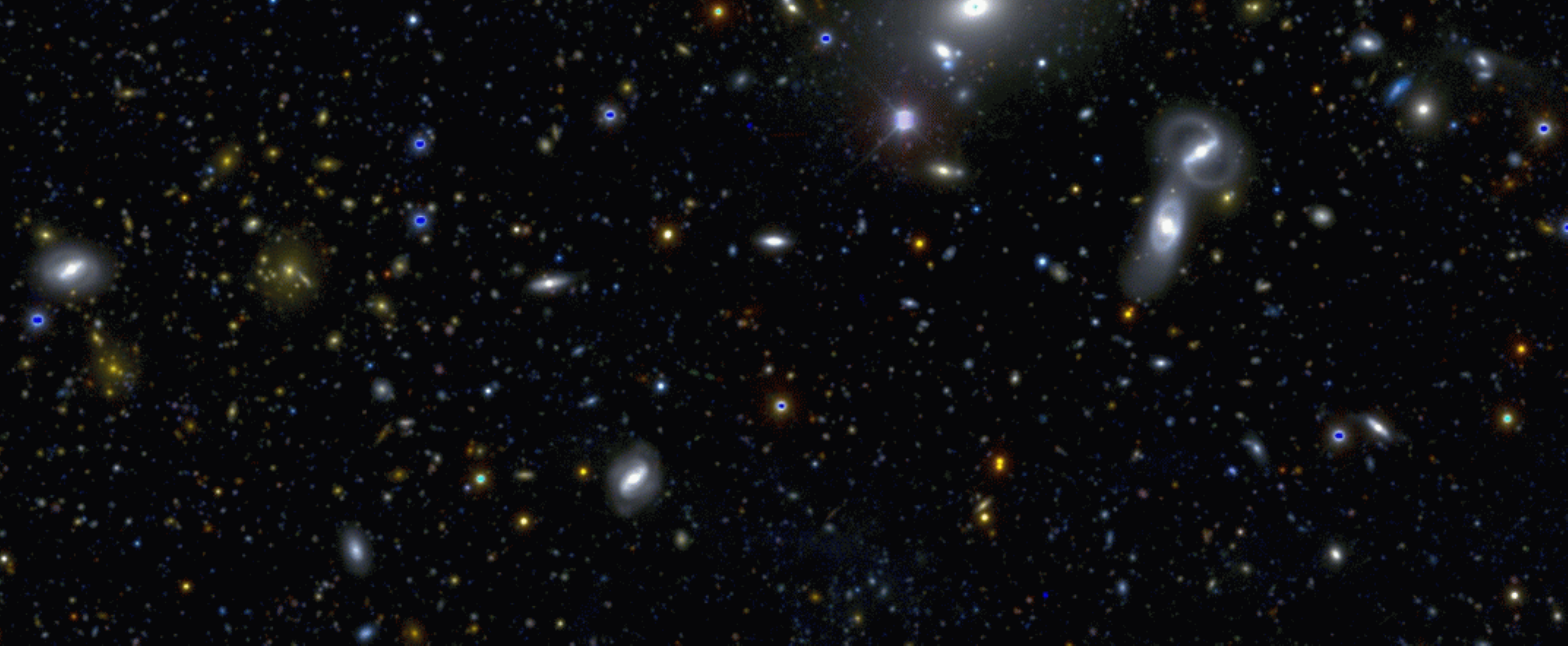}
\caption{A 10\arcmin $\times$ 4\arcmin subsections of coadd tile DES0227-0458 (the center of the tile is at 
Right Ascension 02$^h$27$^m$, Declination -4$^\circ$58\arcmin). The RGB channels of the image 
are the $i+z$ bands, the $r$ band and the $g$ band, respectively. The top panel shows 
typical 2 year wide-area coadd depth. The bottom panel shows increased depth allow in supernova fields after three years. 
Saturated pixels appear as blue in the bottom figure.}\label{fig:COADD}
\end{figure}

At this point our coadded images are complete. Fig. \ref{fig:COADD} shows a 10\arcmin $\times$ 4\arcmin\
rectangle of sky at varying depths. This particular field overlaps the DES SN-X3 SN field and
thus many repeated visits are available to coadd. The top cut combines 4 Final Cut exposures in each 
of the $g$, $r$, $i$ and $z$ bands to produce coadd images with effective exposure time of roughly 360 
seconds in each band. The bottom panel uses the supernova coadds of the same area, includes 10-50 times 
the exposure time, depending on band, and reaches 1-2 magnitudes deeper.

%% file: coadd/coadd-psfex.tex
\subsection{Coadded PSF Extraction}\label{sect:coadd-psfex}

We use \textsc{PSFEx} to produce PSF models for the single-band and detection coadded images using a method similar to 
that presented in Section \ref{sect:psfex}. To produce star catalogs, we run \textsc{SExtractor} with the same 
parameters as in the single epoch run (Appendix \ref{sect:software1}, Table \ref{tab:psfexcat}) except with 
a more conservative \texttt{DETECT\_THRESH} = 5.0. We run \textsc{SExtractor} with identical parameters to those in Table 
\ref{tab:psfex}. We also add parameters \texttt{SAMPLE\_WFLAGMASK= 0x0000} and \texttt{SAMPLE\_IMAFLAGMASK= 0x0000}.

\textsc{PSFEx} does not model PSFs in coadded images as well as in individual exposures. Each coadd tile is a stack of 
dithered individual CCDs from different exposures. The discontinuous boundaries at the edge of the input images lead to 
discontinuities in the PSF that \textsc{PSFEx} does not model. Ultimately, this inaccurate PSF model results in imprecise 
photometry and morphology measurements from the coadded images. This is not a major issue in the detection image, because it is primarily 
used to detect objects and not for precise photometry. However, it limits photometric precision in the single-band coadded images. In areas with heterogeneous
PSFs, photometric precision can be limited to a few percent. In Section \ref{sect:coadd-post}, we describe the simultaneous 
fitting of data from all epochs on the sources from the detection images in individual exposures where the PSF can be modeled accurately \citep{Y1A1}.

%% file: coadd/coadd-cat.tex
\subsection{Coadd Catalogs}\label{sect:coadd-cat}
We use \textsc{SExtractor} produce catalogs from the coadded images using the settings in Appendix \ref{sect:software3}, Table \ref{tab:coadd-sex}.
We use the $r + i + z$ detection image to find and localize sources with a detection threshold of ${\rm S/N} \sim 10$.
We then run \textsc{SExtractor} in dual image mode using the using the $r + i + z$ detection image to define the location of the sources and the single-band images to obtain photometry. 
These catalogs comprise our primary object tables, and we add a series of useful quantities to the objects within them.
The catalogs from the DES Y1+Y2+Y3 are comprised of 400 million objects.

The set of parameters measured for every object includes basic flag and position values 
(\texttt{FLAGS}, \texttt{ALPHAWIN\_J2000\_DET}, \texttt{DELTAWIN\_J2000\_DET}), and calibrated photometry (\texttt{MAG\_AUTO} and \texttt{MAG\_PSF} magnitudes) for every object. 
Caution should be used with the \texttt{MAG\_PSF} quantities due to the heterogeneous nature of the coadd PSF (see Section \ref{sect:coadd-psfex}). 
To mitigate the impact of a poor PSF fitting over the coadded images, we provide a set of weighted-mean quantities.
These weighted-mean quantities are calculated from single-epoch detections within 1\arcsec\ of each object in the coadd detection catalog. 
Two of the most useful weighted-mean quantities are \texttt{WAVG\_MAG\_PSF} and \texttt{WAVG\_SPREAD\_MODEL}.\footnote{\texttt{SPREAD\_MODEL} is a \textsc{SExtractor} parameter derived from the Fisher's linear discriminant between a model of the PSF and an extended source model convolved with the PSF \citep{2012ApJ...757...83D}.}
The individual PSF magnitudes are weighted by the inverse statistical uncertainty (with calibration uncertainty added in quadrature) squared. 
The individual \texttt{SPREAD\_MODEL} are weighted by the inverse \texttt{SPREADERR\_MODEL} squared.
Unfortunately these mean quantities are only available down to the single-epoch detection limit, which is roughly a magnitude shallower than the coadd.
Coadd objects have the \texttt{IMAFLAGS\_ISO} parameter set if they contain pixels that have masked bits set in \emph{all} input images. 
This flag is predominantly set for saturated objects and objects with missing data.

We also run \textsc{SExtractor} separately on the detection image alone to determine our canonical sky position for each 
source. We use the model PSFs from Section \ref{sect:coadd-psfex}. As in Section \ref{sect:fccat}, we convolve our 
images with a $7 \times 7$ pixel Gaussian with a 3 pixel FWHM filter as part of the \textsc{SExtractor} run. Sources in the roughly 30\arcsec\ overlap between 
two tiles (described in Section \ref{sect:coadd-tiles}) can be multiply detected. In this case we only report the detection closest to a tile 
center in the catalogs.



We append several ``value added'' columns to the catalogs.
To account for interstellar extinction from dust, we include $E(B-V)$
values from the \citet[SFD;][]{1998ApJ...500..525S} reddening map at the location of each catalog object.
The $E(B-V)$ values are obtained using a linear interpolation of the originally distributed map in Lambert polar equal-area projection (i.e., \texttt{dust\_getval.c} from Schlegel).
The conversion from $E(B-V)$ to an extinction corrected magnitude is
an active research topic with DES, and we leave it's derivation to
future work \citep{DR1}.
While calculating $E(B-V)$ for the catalog, we calculate Galactic
coordinates, $(l, b)$, which are included in the final catalog.

We assign each source in the catalog a HEALPix pixel \citep{2005ApJ...622..759G} using the \textsc{Healpy} wrapper.\footnote{\url{http://healpix.sourceforge.net}}
This provides simple spatial binning for plotting and statistical angular measurements. We assign each source its pixel value using $\phi = $RA, 
$\theta = 90^\circ - $Dec and \texttt{nside}$ = 2^{14} = 16384$. Each HEALPix pixel has an area of roughly 166 arcsec$^2$.

The above parameters are among the most important in our coadd catalogs, but do not represent an exhaustive list which 
is available on our public data release website.\footnote{\url{https://des.ncsa.illinois.edu/releases}}

%% file: coadd/coadd-post.tex
\subsection{Data Products for Cosmological Analysis}\label{sect:coadd-post}

At this stage, we have produced final images and detection 
catalogs for both the single epoch and coadded data. However, additional data products are required for cosmological analyses. We require depth maps, optimal photometry, 
morphology measurements and photometric redshifts. These data products are not included with DR1, but will be publicly released 
some time after the main DES data release.

We produce spatial depth maps using the \textsc{Mangle} code \citep{2008MNRAS.387.1391S, 2012ascl.soft02005S}. 
\textsc{Mangle}'s basic function is to make regions (masks) as a set of covering, disjoint polygons on a 
spherical surface. These regions carry with them various measures of observing conditions and 
data quality (e.g. FWHM, exposure time, $t_{\rm eff}$ and other 
image characteristics). In our processing, \textsc{Mangle} makes a model for each exposure (working with individual CCD images) 
using its astrometry information to create the basic outline and previously saved outlines of masked (zero weight) 
regions of saturated stars, bleed trails and streaks. For a given tile (Section \ref{sect:coadd-tiles}), 
\textsc{Mangle} uses the outlines of all CCDs in the coadded image to form a set of polygons that represent the final  
coadded image. We use this mask and exposure information to construct maps of depth and related quantities (e.g. 
limiting magnitude, PSF and sky variance) at every point in out survey. These quantities are then linked to the final 
objects catalogs using the object IDs and locations from the coadd catalogs. Our use of \textsc{Mangle} is described in more detail in \citet{Y1A1}. 

We also produce Multi Epoch Data Structures (MEDS) for future analysis \citep{2016MNRAS.460.2245J}. MEDS files include image, PSF and catalog 
information on every object detected in the coadd detection image for each tile. Specifically, they contain 
postage stamps of the science, weight, mask, background and \textsc{SExtractor} segmentation map images for each object 
from both the coadd and all single-epoch images. They also contain models of the local PSFs from \textsc{PSFEx}.
We determine the size of each region from the \textsc{SExtractor} catalog value \texttt{FLUX\_RADIUS} and round this value up to the next integer multiple 
of 16 pixels. Typically, this is 32 or 48 pixels. In addition we pass along photometric zeropoints and astrometry information for each object so that a complete 
analysis can be performed on each object using the MEDS file alone.

We use the output from MEDS to perform multi-epoch, multi-object fitting (MOF) to obtain more robust 
photometric and morphological measurements of every object \citet{Y1A1}. 
MOF uses the PSF and location information from MEDS to fit a stellar 
(unresolved PSF) model and a composite galaxy model to every detected object. MOF also addresses the problem of 
near-neighbor flux contamination by simultaneously fitting groups of sources that have been linked with a 
friends-of-friends algorithm. We use this MOF photometry as our canonical photometry for all cosmological analyses. 
It is superior to single-epoch and coadd photometry for most applications. 
 
We measure galaxy ellipticity and ultimately gravitational lensing shear using the Metacalibration method described in
\citet{2017arXiv170202600H} and \citet{2017arXiv170202601S}. This method simulates a small shear on the images in 
the MEDS postage stamps and measures the derivative of ellipticity with respect to shear. This quantity can in 
turn be robustly related to the underlying gravitational lensing shear.

Finally, we currently derive photometric redshifts for our objects using a template fitting procedure as an afterburner
described in some detail in \citet{2017arXiv170801532H}. This pipeline step is still under active development.

%% file: summ/summ.tex
\section{Conclusion}\label{sect:summ}

This paper has described the pipelines used to process raw DES exposures into science images 
and catalogs. The processes described here were used through the survey's first 
three years and its first public data release.
DES produces multi-band images with an unprecedented combination of depth, 
image quality, and area. The DES Collaboration uses this data to probe the history of cosmic expansion 
and the nature of dark energy using four probes: weak lensing, galaxy clusters, large scale
structure and supernova observations.  The public release of the DES data will only increase 
the breadth of scientific investigation.

The development and operation of the DES science pipelines has been central to the success
of DES, while also serving as a pathfinder for the next generation of ground-based 
imaging experiments -- i.e., the Large Synoptic Survey Telescope's 10-year survey.
The maturation of the DES science pipeline has been driven by two 
needs.  First, the execution of the survey itself, and second, the science needs of the
DES Collaboration.  These have led directly to the split between First Cut and Final Cut and
allows the processing to be focused toward an immediate goal.  In the case of First Cut,
the goal is to feedback data quality to the observing team within 24 hours of an observation. For 
Final Cut the goal is to ensure consistency, uniformity and to maximize the science realizable 
with any one data release.  
By keeping the elements within First Cut and Final Cut as similar as possible, we can also use 
the First Cut 
processing to 1) help inform what updates to codes and/or calibrations are needed, 2) provide
subsequent reprocessing campaigns with data quality assessments to determine which observations 
should be included in release processing campaigns, and 3) enable early testing of algorithmic
improvements at scale.  Moreover, because the pipeline products are similar, First Cut and 
Final Cut products can be combined to enable early scientific analysis of transient phenomena
close in time to when the observations occurred.

Many large astronomical surveys develop their own software suites for data processing. Conversely, 
DES used the \textsc{Astromatic} package (\textsc{SExtractor}, \textsc{PSFEx}, \textsc{SCAMP} 
and \textsc{SWarp}) and \textsc{HOTPANTS} for its core processing tasks. After significant experimenting 
with input parameters and some work with the developer, we found this software performed satisfactorily. 
Having developed our core software early allowed us to focus on DECam-specific development (e.g. the 
sky subtraction and masking algorithms) and cosmology code (e.g. MEDS, MOF and Metacalibration) and 
was essential for developing and end to end survey in a timely manner. 

DES originally planned for yearly internal data releases/realizations.  By the time the Y1 
release occurred it was clear that detailed feedback from the DES Collaboration toward 
continued improvement of the pipelines would not match that cadence.  Thus DES has moved to
bi-annual data releases so that improvements based on the scientific analysis could be 
implemented before the next release was produced.  

Underlying the science algorithm improvements
in the pipelines is a workflow system that tracks detailed 
processing (and reprocessing) provenance, processing meta-data, and data quality assessment with enough 
flexibility to enable integration of changes and new algorithms.
The present DESDM workflow and pipeline system, with a development, release, feedback and reprocessing 
cycle built-in, has served DES well through its internal and public data releases to date.  We look 
forward to carrying out further refinements in the near future to obtain the most out of the
extraordinary DES data set.

%% file: acknowledgements.tex
\acknowledgements

The DESDM team acknowledges support from National Science Foundation
through awards NSF AST 07-15036, NSF AST 08-13543 as well as
significant seed funding provided by the National Center for
Supercomputing Applications and the University of Illinois Department
of Astronomy, the College of Language Arts and Science, and the Vice
Chancellor for Research. DESDM activities in Munich have been
supported by the Ludwig-Maximilians University and the Excellence
Cluster Universe, which is supported by the Deutsche
Forschungsgemeinschaft (DFG). 

This work made use of the Illinois Campus Cluster, a computing resource 
that is operated by the Illinois Campus Cluster Program (ICCP) in conjunction 
with the National Center for Supercomputing Applications (NCSA) and which is 
supported by funds from the University of Illinois at Urbana-Champaign.

This research is part of the Blue Waters sustained-petascale computing project, 
which is supported by the National Science Foundation (awards OCI-0725070 and 
ACI-1238993) and the state of Illinois. Blue Waters is a joint effort of the 
University of Illinois at Urbana-Champaign and its National Center for 
Supercomputing Applications.

Funding for the DES Projects has been provided by the U.S. Department of Energy, the U.S. National Science Foundation, the Ministry of Science and Education of Spain, 
the Science and Technology Facilities Council of the United Kingdom, the Higher Education Funding Council for England, the National Center for Supercomputing 
Applications at the University of Illinois at Urbana-Champaign, the Kavli Institute of Cosmological Physics at the University of Chicago, 
the Center for Cosmology and Astro-Particle Physics at the Ohio State University,
the Mitchell Institute for Fundamental Physics and Astronomy at Texas A\&M University, Financiadora de Estudos e Projetos, 
Funda{\c c}{\~a}o Carlos Chagas Filho de Amparo {\`a} Pesquisa do Estado do Rio de Janeiro, Conselho Nacional de Desenvolvimento Cient{\'i}fico e Tecnol{\'o}gico and 
the Minist{\'e}rio da Ci{\^e}ncia, Tecnologia e Inova{\c c}{\~a}o, the Deutsche Forschungsgemeinschaft and the Collaborating Institutions in the Dark Energy Survey. 

The Collaborating Institutions are Argonne National Laboratory, the University of California at Santa Cruz, the University of Cambridge, Centro de Investigaciones Energ{\'e}ticas, 
Medioambientales y Tecnol{\'o}gicas-Madrid, the University of Chicago, University College London, the DES-Brazil Consortium, the University of Edinburgh, 
the Eidgen{\"o}ssische Technische Hochschule (ETH) Z{\"u}rich, 
Fermi National Accelerator Laboratory, the University of Illinois at Urbana-Champaign, the Institut de Ci{\`e}ncies de l'Espai (IEEC/CSIC), 
the Institut de F{\'i}sica d'Altes Energies, Lawrence Berkeley National Laboratory, the Ludwig-Maximilians Universit{\"a}t M{\"u}nchen and the associated Excellence Cluster Universe, 
the University of Michigan, the National Optical Astronomy Observatory, the University of Nottingham, The Ohio State University, the University of Pennsylvania, the University of Portsmouth, 
SLAC National Accelerator Laboratory, Stanford University, the University of Sussex, Texas A\&M University, and the OzDES Membership Consortium.

Based in part on observations at Cerro Tololo Inter-American Observatory, National Optical Astronomy Observatory, which is operated by the Association of 
Universities for Research in Astronomy (AURA) under a cooperative agreement with the National Science Foundation.

The DES data management system is supported by the National Science Foundation under Grant Numbers AST-1138766 and AST-1536171.
The DES participants from Spanish institutions are partially supported by MINECO under grants AYA2015-71825, ESP2015-66861, FPA2015-68048, SEV-2016-0588, SEV-2016-0597, and MDM-2015-0509, 
some of which include ERDF funds from the European Union. IFAE is partially funded by the CERCA program of the Generalitat de Catalunya.
Research leading to these results has received funding from the European Research
Council under the European Union's Seventh Framework Program (FP7/2007-2013) including ERC grant agreements 240672, 291329, and 306478.
We  acknowledge support from the Australian Research Council Centre of Excellence for All-sky Astrophysics (CAASTRO), through project number CE110001020.

This manuscript has been authored by Fermi Research Alliance, LLC under Contract No. DE-AC02-07CH11359 with the U.S. Department of Energy, Office of Science, Office of High Energy Physics. The United States Government retains and the publisher, by accepting the article for publication, acknowledges that the United States Government retains a non-exclusive, paid-up, irrevocable, world-wide license to publish or reproduce the published form of this manuscript, or allow others to do so, for United States Government purposes.

%% file: appendix/flags.tex
\section{Appendix A: Input File and Flag Descriptions}\label{sect:flags}

In Section \ref{sect:preprocess}, we describe input files and two sets of flags which we 
tabulate below. Table \ref{tab:files} describes the input files we use in Preprocessing, and
Table \ref{tab:links} gives links to example files.
Table \ref{tab:bpm} shows the Bad Pixel Mask (BPM) flags. These 
flags are applied to BPM images that
are generated once for each set of calibration images, which are in turn 
remade every time the camera warms up (typically twice per year).

\begin{table}
\centering
\begin{tabular}{cccc}
        \hline
Name  &  Description  &  Update Frequency &  Used By  \\
        \hline
Header Update   & GainAB, ReadnoiseAB, SaturateAB amp keywords &  Survey &   Crosstalk  \\
Crosstalk Matrix  &  Matrix of Crosstalk correction between amps/CCDs  & Survey  &  Crosstalk  \\
Brighter-Fatter Correction  &  Matrix for PSF deconvolution for brighter-fatter effect  &  Survey  &   \textsc{Pixcorrect}  \\
Nonlinearity LUT  &  Look up table to correct for Non-linearity  &  Survey  &  \textsc{Pixcorrect}  \\
BPM  &  Bad Pixel Mask  & Epoch &  \textsc{Pixcorrect}  \\
Dome Flat  &  Image of Blanco white spot with DECam  &  Nightly &  \textsc{Pixcorrect}  \\
Dome Superflat  &  Average of 50-100 Dome Flats  & Epoch &  \textsc{Pixcorrect}  \\
Bias  &  zero second exposure  &  Nightly &  \textsc{Pixcorrect} \\
Biascor Supercal  &  Average of 50-100 Bias frames  &  Epoch &  \textsc{Pixcorrect} \\
TPV Distortion  &  Estimate of astrometric distortion of each CCD  &  Epoch &  \textsc{SCAMP} \\
PCA Binned Sky &  128x128 binned sky background PCA components  &  Epoch & Sky Subtraction  \\
PCA Sky Template  &  Full resolution sky background PCA components &  Epoch &  Sky Subtraction  \\
Starflat  &  Multiplicative, large scale (> arcmin) DECam response &  Epoch &  Sky Subtraction \\
        \hline
\end{tabular}
\caption{\rm{Files used by Preprocessing routines. We present them in the order they are used by the pipeline.}}\label{tab:files}
\end{table}

\begin{table}
\centering
\begin{tabular}{cc}
        \hline
Name  &  File Path \\
        \hline
Header Update   & \texttt{config/20170531/finalcut/20170531\_DES\_header\_update.20170329} \\
Crosstalk Matrix  & \texttt{cal/xtalk/20130606/DECam\_20130606.xtalk} \\
Brighter Fatter Correction  & \texttt{cal/bf/20170406-r2959/p01/D\_n20170406\_r2959p01\_bf.fits} \\
Nonlinearity LUT  & \texttt{cal/lintable/20170314/lin\_tbl\_v0.7.fits} \\
BPM  & \texttt{cal/bpm/20170201t0213-r2927/p01/} \\
Dome Flat  & \texttt{precal/20150212-r1402/p01/dflatcor/} \\
Dome Superflat  & \texttt{supercal/Y4N/20170201t0213-r2922/p01/norm-dflatcor/} \\
Bias  & \texttt{precal/Y4N/20170208-r2923/p01/biascor/} \\
Biascor Supercal  & \texttt{supercal/Y4N/20170201t0213-r2922/p01/biascor/} \\
TPV Distortion  & \texttt{config/20170531/finalcut/20170531\_decam\_pvmodel\_Y4E2\_i.ahead} \\
PCA Binned Sky & \texttt{cal/skytemp/20170103t0315-r2932/p01/binned-fp/ } \\
PCA Sky Template  & \texttt{cal/skytemp/20170103t0315-r2932/p01/tmpl/} \\
Starflat  & \texttt{cal/starflat/20170103t0315-r2933/p01/} \\
        \hline
\end{tabular}
\caption{\rm{Links to samples of files used by Preprocessing routines. These links are all prefaced with 
\texttt{http://data.darkenergysurvey.org/aux/sample/}. 
In cases where only a directory is given, the directory contains many examples of the file in question, generally one per CCD and/or band.}}\label{tab:links}
\end{table}

\begin{table}
\centering
\begin{tabular}{llll}
        \hline
BPM Name & Value & Hex & Description \\
        \hline
\texttt{BPMDEF\_FLAT\_MIN}  & \texttt{1}     & \texttt{0x0001} & Flat pixel $<$ 50\% mean\\
\texttt{BPMDEF\_FLAT\_MAX}  & \texttt{2}     & \texttt{0x0002} & Flat pixel $>$ 150\% mean\\
\texttt{BPMDEF\_BIAS\_HOT}  & \texttt{8}     & \texttt{0x0008} & Bias pixel $>$ 500 counts\\
\texttt{BPMDEF\_BIAS\_WARM} & \texttt{16}    & \texttt{0x0010} & Bias pixel $>$ 20 counts\\
\texttt{BPMDEF\_BIAS\_COL}  & \texttt{64}    & \texttt{0x0040} & Downstream (in readout order) from hot pixel\\
\texttt{BPMDEF\_EDGE}       & \texttt{128}   & \texttt{0x0080} & Nearest 15 pixels to the CCD edge\\
\texttt{BPMDEF\_SUSPECT}    & \texttt{512}   & \texttt{0x0200} & Regions with small, non-quantifiable errors (See text) \\
\texttt{BPMDEF\_FUNKY\_COL} & \texttt{1024}  & \texttt{0x0400} & Columns with anomalous sky values \\
\texttt{BPMDEF\_WACKY\_PIX} & \texttt{2048}  & \texttt{0x0800} & Areas with anomalous sky values \\
\texttt{BPMDEF\_BADAMP}     & \texttt{4096}  & \texttt{0x1000} & Unusable amplifier (right amp of DECam CCD 31)\\
\texttt{BPMDEF\_NEAREDGE}   & \texttt{8192}  & \texttt{0x2000} & 25 edge columns unreliable on 1-2\% level \\
\texttt{BPMDEF\_TAPEBUMP}   & \texttt{16384} & \texttt{0x4000} & Each CCD has 6 tape bumps (See text)\\
        \hline
\end{tabular}
\caption{\rm{Bad Pixel Mask (BPM) bit descriptions. Values are listed as conventional integers and in hexadecimal.}}\label{tab:bpm}
\end{table}

Table \ref{tab:bitmask} shows the bitmask flags. These are generated for 
each individual exposure. Some flags are derived from the BPM and are 
constant for every exposure in an epoch. Others mark the saturated regions, 
cosmic rays, streaks and other defects in individual exposures.

\begin{table*}
\centering
\begin{tabular}{llll}
        \hline
Bitmask Name & Value & Hex & Description \\
        \hline
\texttt{BADPIX\_BPM}       & \texttt{1}     & \texttt{0x0001} & BPM value \texttt{0x0c5b} (see exceptions in text)\\
\texttt{BADPIX\_SATURATE}  & \texttt{2}     & \texttt{0x0002} & Pixel above amplifier's saturation value in debiased image\\
\texttt{BADPIX\_INTERP}    & \texttt{4}     & \texttt{0x0004} & Pixel whose value was interpolated from neighbors\\
\texttt{BADPIX\_BADAMP}    & \texttt{8}     & \texttt{0x0008} & Identical to \texttt{BPMDEF\_BADAMP} (BPM=\texttt{0x1000})\\
\texttt{BADPIX\_CRAY}      & \texttt{16}    & \texttt{0x0010} & Pixel contaminated by a cosmic ray\\
\texttt{BADPIX\_STAR}      & \texttt{32}    & \texttt{0x0020} & Circular mask around bright stars \\
\texttt{BADPIX\_TRAIL}     & \texttt{64}    & \texttt{0x0040} & Bleed trails form in column with saturated pixels\\
\texttt{BADPIX\_EDGEBLEED} & \texttt{128}   & \texttt{0x0080} & Near CCD edges, bleed trails can spread to adjacent pixels\\
\texttt{BADPIX\_SSXTALK}   & \texttt{256}   & \texttt{0x0100} & Cross talk from saturated pixels is uncertain (though typically small)\\
\texttt{BADPIX\_EDGE}      & \texttt{512}   & \texttt{0x0200} & Identical to \texttt{BPMDEF\_EDGE} (BPM= \texttt{0x0080})\\
\texttt{BADPIX\_STREAK}    & \texttt{1024}  & \texttt{0x0400} & Streak detected\\
\texttt{BADPIX\_SUSPECT}   & \texttt{2048}  & \texttt{0x0800} & Identical to \texttt{BPMDEF\_SUSPECT} (BPM= \texttt{0x0200})\\
\texttt{BADPIX\_FIXED}     & \texttt{4096}  & \texttt{0x1000} & Fixed column with \texttt{BPMDEF\_BIAS\_COL} or \texttt{BPMDEF\_FUNKY\_COL}\\
\texttt{BADPIX\_NEAREDGE}  & \texttt{8129}  & \texttt{0x2000} & Identical to \texttt{BPMDEF\_NEAREDGE} (BPM=\texttt{0x2000})\\
\texttt{BADPIX\_TAPEBUMP}  & \texttt{16384} & \texttt{0x4000} & Identical to \texttt{BPMDEF\_TAPEBUMP} (BPM=\texttt{0x4000})\\
        \hline
\end{tabular}
\caption{\rm{Bitmask descriptions for individual flag images. Values are listed as conventional integers and in hexadecimal.}}\label{tab:bitmask}
\end{table*}

%% file: appendix/software1.tex
\section{Appendix B: Software Settings for the First Cut and Final Cut Pipelines}\label{sect:software1}

Tables \ref{tab:scampsex} through \ref{tab:firstcutsex} are used in First Cut (Section \ref{sect:firstcut}) and Final Cut (Section \ref{sect:finalcut}) 
processing. Tables \ref{tab:scampsex} and \ref{tab:scamp} describe the \textsc{SExtractor} settings used to produce a star catalog for 
\textsc{SCAMP} and the corresponding \textsc{SCAMP} settings (see Section \ref{sect:scamp}). Similarly, Tables \ref{tab:psfexcat} and \ref{tab:psfex} 
describe the \textsc{SExtractor} settings used to produce a star catalog for \textsc{PSFEx} and the corresponding \textsc{PSFEx} settings (see Section 
\ref{sect:psfex}). Finally, the \textsc{SExtractor} settings for our First Cut and Final Cut source catalogs are listed in Table \ref{tab:firstcutsex}.

\begin{table}
\centering
\begin{tabular}{ll}
        \hline
Parameter & Value \\
        \hline
\texttt{DETECT\_MINAREA} & $5$ \\
\texttt{DETECT\_THRESH}  & $10$ \\
\texttt{ANALYSIS\_THRESH} & $1.5$ \\
\texttt{WEIGHT\_THRESH}  & $10^{-5}$ \\
\texttt{FILTER}  &        \texttt{Y} \\
\texttt{FILTER\_NAME} & \textit{See text}\\
\texttt{DEBLEND\_NTHRESH} & $32$ \\
\texttt{DEBLEND\_MINCONT} & $0.1$ \\
\texttt{CLEAN}           & \texttt{Y} \\
\texttt{CLEAN\_PARAM}     & $1.0$\\
\texttt{MASK\_TYPE}       & \texttt{CORRECT}  \\
\texttt{INTERP\_TYPE}     & \texttt{ALL}\\
\texttt{INTERP\_MAXXLAG}   &  $4$\\
\texttt{INTERP\_MAXYLAG}   & $4$\\
\texttt{SEEING\_FWHM}    & 1.2$"$\\
\texttt{STARNNW\_NAME}    & \texttt{DEFAULT}\\
\texttt{RESCALE\_WEIGHTS} & \texttt{N}\\
\texttt{BACK\_SIZE}       & $256$\\
\texttt{BACK\_FILTERSIZE} & $3$\\
\texttt{BACKPHOTO\_TYPE}       & \texttt{GLOBAL}\\
\texttt{PSF\_NMAX}       & $1$  \\
\texttt{BACK\_TYPE}       & \texttt{AUTO}\\
        \hline
\end{tabular}
\caption{\rm{\textsc{SExtractor} parameters for \textsc{SCAMP} analysis.}}\label{tab:scampsex}
\end{table}

\begin{table}

\centering
\begin{tabular}{ll}
        \hline
Parameter & Value \\
        \hline
\texttt{FGROUP\_RADIUS} & \texttt{3.0}\\
\texttt{ASTREF\_CATALOG}  &  \texttt{2MASS} \\
\texttt{ASTREF\_BAND}  &  \texttt{DEFAULT} \\
\texttt{ASTREFCENT\_KEYS}  &  \texttt{X\_WORLD,Y\_WORLD} \\
\texttt{ASTREFERR\_KEYS}  &  \texttt{ERRA\_WORLD, ERRTHETA\_WORLD} \\
\texttt{ASTREFMAG\_KEY}  &  \texttt{MAG} \\
\texttt{MATCH}  &  \texttt{Y} \\
\texttt{MATCH\_NMAX}  &  $0$ \\
\texttt{PIXSCALE\_MAXERR}  &  $1.2$ \\
\texttt{POSANGLE\_MAXERR}  &  $5.0$ \\
\texttt{POSITION\_MAXERR}  &  $10.0$ \\
\texttt{MATCH\_RESOL}  &  $0$ \\
\texttt{MATCH\_FLIPPED}  &  \texttt{N} \\
\texttt{MOSAIC\_TYPE}  &  \texttt{SAME\_CRVAL} \\
\texttt{FIXFOCALPLANE\_NMIN}  &  $1$ \\
\texttt{CROSSID\_RADIUS}  &  $2.0$ \\
\texttt{SOLVE\_ASTROM}  &  \texttt{Y} \\
\texttt{ASTRINSTRU\_KEY}  &  \texttt{FILTER,QRUNID} \\
\texttt{STABILITY\_TYPE}  &  \texttt{INSTRUMENT} \\
\texttt{CENTROID\_KEYS}  &  \texttt{XWIN\_IMAGE, YWIN\_IMAGE} \\
\texttt{CENTROIDERR\_KEYS}  &  \texttt{ERRAWIN\_IMAGE, ERRBWIN\_IMAGE,}\\
                            & \texttt{ERRTHETAWIN\_IMAGE} \\
\texttt{DISTORT\_KEYS}  &  \texttt{XWIN\_IMAGE, YWIN\_IMAGE} \\
\texttt{DISTORT\_GROUPS}  &  $1, 1$ \\
\texttt{DISTORT\_DEGREES}  &  $3$ \\
\texttt{FOCDISTORT\_DEGREE}  &  $2$ \\
\texttt{ASTREF\_WEIGHT}  &  $1.0$ \\
\texttt{ASTRCLIP\_NSIGMA}  &  $3.0$ \\
\texttt{CORRECT\_COLOURSHIFTS}  &  \texttt{N} \\
\texttt{SOLVE\_PHOTOM}  &  \texttt{Y} \\
\texttt{MAGZERO\_OUT}  &  $0.0$ \\
\texttt{MAGZERO\_INTERR}  &  $0.01$ \\
\texttt{MAGZERO\_REFERR}  &  $0.03$ \\
\texttt{ASTREFMAG\_LIMITS}  &  $-99, 17$ \\
\texttt{PHOTCLIP\_NSIGMA}  &  $3.0$ \\
\texttt{SN\_THRESHOLDS}  &  $10.0, 100.0$ \\
\texttt{FWHM\_THRESHOLDS}  &  $0.0, 50.0$ \\
\texttt{FLAGS\_MASK}  &  $0x00f0$ \\
\texttt{WEIGHTFLAGS\_MASK}  &  $0x00ff$ \\
\texttt{IMAFLAGS\_MASK}  &  $0x0000$ \\
        \hline
\end{tabular}
\caption{\rm{First Cut \textsc{SCAMP} settings.}}\label{tab:scamp}
\end{table}

\begin{table}
\centering
\begin{tabular}{ll}
        \hline
Parameter & Value \\
        \hline
\texttt{DETECT\_MINAREA} & $3$ \\
\texttt{DETECT\_THRESH}  & $3.0$ \\
\texttt{ANALYSIS\_THRESH} & $1.5$ \\
\texttt{FILTER}  &         \texttt{Y} \\
\texttt{FILTER\_NAME} & all-ground\\
\texttt{DEBLEND\_NTHRESH} & $32$ \\
\texttt{DEBLEND\_MINCONT} & $0.005$ \\
\texttt{CLEAN}           & \texttt{Y} \\
\texttt{CLEAN\_PARAM}     & $1.0$\\
\texttt{MASK\_TYPE}       & \texttt{CORRECT}  \\
\texttt{PSF\_NMAX}       & $1$  \\
\texttt{STARNNW\_NAME}    & \texttt{DEFAULT}\\
\texttt{RESCALE\_WEIGHTS} & \texttt{N}\\
\texttt{BACK\_SIZE}       & $256$\\
\texttt{BACK\_FILTERSIZE} & $3$\\
\texttt{BACKPHOTO\_TYPE}       & \texttt{GLOBAL}\\
\texttt{RESCALE\_WEIGHTS}       & \texttt{N}\\
\texttt{BACK\_TYPE}       & \texttt{AUTO}\\
        \hline
\end{tabular}
\caption{\rm{\textsc{SExtractor} parameters used to create the catalog for \textsc{PSFEx} analysis.}}\label{tab:psfexcat}
\end{table}

\begin{table}
\centering
\begin{tabular}{ll}
        \hline
Parameter & Value \\
        \hline
\texttt{BASIS\_TYPE}  &  \texttt{PIXEL\_AUTO} \\
\texttt{BASIS\_NUMBER}  &  $20$ \\
\texttt{BASIS\_SCALE}  &  $1.0$ \\
\texttt{NEWBASIS\_TYPE}  &  \texttt{NONE} \\
\texttt{NEWBASIS\_NUMBER}  &  $8$ \\
\texttt{PSF\_SAMPLING}  &  $0.0$ \\
\texttt{PSF\_PIXELSIZE}  &  $1.0$ \\
\texttt{PSF\_ACCURACY}  &  $0.01$ \\
\texttt{PSF\_SIZE}  &  $25,25$ \\
\texttt{CENTER\_KEYS}  &  \texttt{X\_IMAGE,Y\_IMAGE} \\
\texttt{PSF\_RECENTER}  &  \texttt{N} \\
\texttt{PHOTFLUX\_KEY}  &  \texttt{FLUX\_APER(7)} \\
\texttt{PHOTFLUXERR\_KEY}  &  \texttt{FLUXERR\_APER(7)} \\
\texttt{MEF\_TYPE}  &  \texttt{INDEPENDENT} \\
\texttt{PSFVAR\_KEYS}  &  \texttt{X\_IMAGE,Y\_IMAGE} \\
\texttt{PSFVAR\_GROUPS}  &  $1,1$ \\
\texttt{PSFVAR\_DEGREES}  &  $2$ \\
\texttt{PSFVAR\_NSNAP}  &  $9$ \\
\texttt{HIDDENMEF\_TYPE}  &  \texttt{COMMON} \\
\texttt{STABILITY\_TYPE}  &  \texttt{PRE-DISTORTED} \\
\texttt{SAMPLE\_AUTOSELECT}  &  \texttt{Y} \\
\texttt{SAMPLEVAR\_TYPE}  &  \texttt{SEEING} \\
\texttt{SAMPLE\_FWHMRANGE}  &  $2.0,30.0$ \\
\texttt{SAMPLE\_VARIABILITY}  &  $0.2$ \\
\texttt{SAMPLE\_MINSN}  &  $20$ \\
\texttt{SAMPLE\_MAXELLIP}  &  $0.3$ \\
\texttt{SAMPLE\_FLAGMASK}  &  $0x00fe$ \\
\texttt{BADPIXEL\_FILTER}  &  \texttt{N} \\
\texttt{BADPIXEL\_NMAX}  &  $0$ \\
\texttt{HOMOBASIS\_TYPE}  &  \texttt{None} \\
\texttt{HOMOBASIS\_NUMBER}  &  $10$ \\
\texttt{HOMOBASIS\_SCALE}  &  $1.0$ \\
\texttt{HOMOPSF\_PARAMS}  &  $2.0,3.0$ \\
\texttt{CHECKPLOT\_DEV}  &  \texttt{NULL} \\
\texttt{CHECKPLOT\_RES}  &  $0$ \\
\texttt{CHECKPLOT\_ANTIALIAS}  &  \texttt{Y} \\
        \hline
\end{tabular}
\caption{\rm{\textsc{PSFEx} parameters used for measuring and modeling DES PSFs.}}\label{tab:psfex}
\end{table}

\begin{table}
\centering
\begin{tabular}{ll}
        \hline
Parameter & Value \\
        \hline
\texttt{DETECT\_MINAREA} & $6$ \\
\texttt{DETECT\_THRESH}  & $1.5$ \\
\texttt{ANALYSIS\_THRESH} & $1.5$ \\
\texttt{FILTER}  &         \texttt{Y} \\
\texttt{FILTER\_NAME} & all-ground\\
\texttt{DEBLEND\_NTHRESH} & $32$ \\
\texttt{DEBLEND\_MINCONT} & $0.005$ \\
\texttt{CLEAN}           & \texttt{Y} \\
\texttt{CLEAN\_PARAM}     & $1.0$\\
\texttt{MASK\_TYPE}       & \texttt{CORRECT}  \\
\texttt{PSF\_NMAX}       & $1$  \\
\texttt{STARNNW\_NAME}    & \texttt{DEFAULT}\\
\texttt{RESCALE\_WEIGHTS} & \texttt{N}\\
\texttt{BACK\_SIZE}       & $256$\\
\texttt{BACK\_FILTERSIZE} & $3$\\
\texttt{BACKPHOTO\_TYPE}       & \texttt{GLOBAL}\\
\texttt{BACK\_TYPE}       & \texttt{AUTO}\\
\texttt{INTERP\_TYPE}     & \texttt{VAR\_ONLY}\\
\texttt{INTERP\_MAXXLAG}   & $4$\\
\texttt{INTERP\_MAXYLAG}   & $4$\\
        \hline
\end{tabular}
\caption{\rm{\textsc{SExtractor} parameters for DES First Cut catalog production.}}\label{tab:firstcutsex}
\end{table}

%% file: appendix/software2.tex
\section{Appendix C: Software Settings for the Supernova Pipeline}\label{sect:software2}

The DES Supernova pipeline (Section \ref{sect:supernova}) has settings independent of the main pipeline that we describe in 
Tables \ref{tab:snscampsex} through \ref{tab:snsex}. Tables \ref{tab:snscampsex} and \ref{tab:snscamp} describe the \textsc{SCAMP} 
settings used to perform the astrometric refinement used for DES difference imaging. The settings in Tables \ref{tab:snpsfexcat} 
and \ref{tab:snpsfex} are used to model the PSF with PSFEx. The supernova pipeline then astrometrically resamples the First Cut-processed 
image using \textsc{SWarp} and the settings in Table \ref{tab:SWarp}. We produce a difference image using \textsc{HOTPANTS} and the settings 
in Table \ref{tab:hotpants}, and use \textsc{SExtractor} and the settings in \ref{tab:snsex} to produce an initial catalog of potential 
transients. 

\begin{table}
\centering
\begin{tabular}{ll}
        \hline
Parameter & Value \\
        \hline
\texttt{DETECT\_MINAREA} & $3$\\
\texttt{DETECT\_THRESH} & $5$\\
\texttt{ANALYSIS\_THRESH} & \texttt{5}\\
\texttt{FILTER} & \texttt{Y}\\
\texttt{FILTER\_NAME} & all-ground\\
\texttt{DEBLEND\_MINCONT} & $0.05$\\
\texttt{WEIGHT\_TYPE} & \texttt{MAP\_WEIGHT}\\
\texttt{WEIGHT\_GAIN} & \texttt{Y}\\
\texttt{WEIGHT\_THRESH} & $1.e-10$\\
\texttt{PHOT\_APERTURES} & $10$\\
\texttt{PHOT\_AUTOPARAMS} & $3.5,3.5$\\
\texttt{SATUR\_LEVEL} & $15000.0$\\
\texttt{GAIN} & \texttt{1.0}\\
\texttt{INTERP\_TYPE} & \texttt{NONE}\\
        \hline
\end{tabular}
\caption{\rm{\textsc{SExtractor} settings for catalogs used in the astrometric refinement in the supernova pipeline.}}\label{tab:snscampsex}
\end{table}

\begin{table}
\centering
\begin{tabular}{ll}
        \hline
Parameter & Value \\
        \hline
\texttt{FGROUP\_RADIUS} & $1.0$\\
\texttt{ASTREF\_CATALOG} & See text\\
\texttt{ASTREF\_BAND} & \texttt{DEFAULT}\\
\texttt{ASTREFCENT\_KEYS} & \texttt{X\_WORLD,Y\_WORLD}\\
\texttt{ASTREFERR\_KEYS} & \texttt{ERRA\_WORLD,ERRB\_WORLD}\\
                         & \texttt{ERRTHETA\_WORLD}\\
\texttt{ASTREFMAG\_KEY} & \texttt{MAG}\\
\texttt{MATCH} & \texttt{N}\\
\texttt{PIXSCALE\_MAXERR} & $1.1$\\
\texttt{POSANGLE\_MAXERR} & $1.0$\\
\texttt{POSITION\_MAXERR} & $0.2$\\
\texttt{MATCH\_RESOL} & $0$\\
\texttt{MATCH\_FLIPPED} & \texttt{N}\\
\texttt{MOSAIC\_TYPE} & \texttt{UNCHANGED}\\
\texttt{FIXFOCALPLANE\_NMIN} & $1$\\
\texttt{CROSSID\_RADIUS} & $5.0$\\
\texttt{SOLVE\_ASTROM} & \texttt{Y}\\
\texttt{PROJECTION\_TYPE} & \texttt{TAN}\\
\texttt{ASTRINSTRU\_KEY} & \texttt{FILTER}\\
\texttt{STABILITY\_TYPE} & \texttt{INSTRUMENT}\\
\texttt{CENTROID\_KEYS} & \texttt{XWIN\_IMAGE,YWIN\_IMAGE}\\
\texttt{CENTROIDERR\_KEYS} & \texttt{ERRAWIN\_IMAGE,ERRBWIN\_IMAGE}\\
                           & \texttt{ERRTHETAWIN\_IMAGE}\\
\texttt{DISTORT\_KEYS} & \texttt{XWIN\_IMAGE,YWIN\_IMAGE}\\
\texttt{DISTORT\_GROUPS} & $1,1$\\
\texttt{DISTORT\_DEGREES} & $1$\\
\texttt{FOCDISTORT\_DEGREE} & $1$\\
\texttt{ASTREF\_WEIGHT} & $1.0$\\
\texttt{ASTRACCURACY\_TYPE} & \texttt{SIGMA-PIXEL}\\
\texttt{ASTRACCURACY\_KEY} & \texttt{ASTRACCU}\\
\texttt{ASTR\_ACCURACY} & $0.01$\\
\texttt{ASTRCLIP\_NSIGMA} & $3.0$\\
\texttt{CORRECT\_COLOURSHIFTS} & \texttt{N}\\
\texttt{SOLVE\_PHOTOM} & \texttt{Y}\\
\texttt{MAGZERO\_OUT} & $0.0$\\
\texttt{MAGZERO\_INTERR} & $0.01$\\
\texttt{MAGZERO\_REFERR} & $0.03$\\
\texttt{ASTREFMAG\_LIMITS} & $-99.0, 99.0$\\
\texttt{PHOTCLIP\_NSIGMA} & $3.0$\\
\texttt{PHOT\_ACCURACY} & $10^{-3}$\\
\texttt{SN\_THRESHOLDS} & $10.0, 100.0$\\
\texttt{FWHM\_THRESHOLDS} & $1.0, 20.0$\\
\texttt{ELLIPTICITY\_MAX} & $0.4$\\
\texttt{FLAGS\_MASK} & $0x00ff$\\
\texttt{WEIGHTFLAGS\_MASK} & $0x00ff$\\
\texttt{IMAFLAGS\_MASK} & $0x000$\\
        \hline
\end{tabular}
\caption{\rm{Our \textsc{SCAMP} settings for astrometric refinement in the supernova pipeline.}}\label{tab:snscamp}
\end{table}

\begin{table}
\centering
\begin{tabular}{cc}
        \hline
Parameter & Value \\
        \hline
\texttt{ASSOCCOORD\_TYPE} & \texttt{WORLD}\\
\texttt{ASSOC\_RADIUS} & $2.0$\\
\texttt{ASSOC\_TYPE} & \texttt{NEAREST}\\
\texttt{ASSOCSELEC\_TYPE} & \texttt{MATCHED}\\
\texttt{DETECT\_MINAREA} & $3$\\
\texttt{DETECT\_THRESH} & $5$\\
\texttt{ANALYSIS\_THRESH} & $5$\\
\texttt{FILTER} & \texttt{Y}\\
\texttt{FILTER\_NAME} & all-ground\\
\texttt{DEBLEND\_MINCONT} & $0.05$\\
\texttt{WEIGHT\_TYPE} & \texttt{MAP\_WEIGHT}\\
\texttt{WEIGHT\_GAIN} & \texttt{Y}\\
\texttt{WEIGHT\_THRESH} & $10^{-10}$\\
\texttt{PHOT\_APERTURES} & $10$\\
\texttt{PHOT\_AUTOPARAMS} & $3.5, 3.5$\\
\texttt{SATUR\_LEVEL} & $15000.0$\\
       \hline
\end{tabular}
\caption{\rm{\textsc{SExtractor} parameters used to create catalogs for supernova \textsc{PSFEx} analysis.}}\label{tab:snpsfexcat}
\end{table}

\begin{table}
\centering
\begin{tabular}{cc}
        \hline
Parameter & Value \\
        \hline
\texttt{BASIS\_TYPE} & \texttt{PIXEL\_AUTO}\\
\texttt{BASIS\_NUMBER} & $16$\\
\texttt{PSF\_SAMPLING} & $0.7$\\
\texttt{PSF\_ACCURACY} & $0.01$\\
\texttt{PSF\_SIZE} & $51, 51$\\
\texttt{PSF\_RECENTER} & \texttt{Y}\\
\texttt{PSFVAR\_KEYS} & \texttt{X\_IMAGE,Y\_IMAGE}\\
\texttt{PSFVAR\_GROUPS} & $1, 1$\\
\texttt{PSFVAR\_DEGREES} & $1$\\
\texttt{SAMPLE\_AUTOSELECT} & \texttt{Y}\\
\texttt{SAMPLEVAR\_TYPE} & \texttt{SEEING}\\
\texttt{SAMPLE\_FWHMRANGE} & $3.0, 20.0$\\
\texttt{SAMPLE\_VARIABILITY} & $0.2$\\
\texttt{SAMPLE\_MINSN} & $20$\\
\texttt{SAMPLE\_MAXELLIP} & $0.3$\\
\texttt{CHECKPLOT\_DEV} & \texttt{NULL}\\
\texttt{CHECKPLOT\_TYPE} & \texttt{NONE}\\
\texttt{CHECKPLOT\_NAME} & \texttt{fwhm,ellipticity}\\
\texttt{CHECKIMAGE\_TYPE} & \texttt{NONE}\\
        \hline
\end{tabular}
\caption{\rm{\textsc{PSFEx} parameters used for measuring and modeling DES PSFs in the supernova fields.}}\label{tab:snpsfex}
\end{table}

\begin{table}
\centering
\begin{tabular}{ll}
        \hline
Parameter & Value \\
        \hline
\texttt{COMBINE} & \texttt{Y} \\
\texttt{COMBINE\_TYPE} & \texttt{AVERAGE} \\
\texttt{BLANK\_BADPIXELS} & \texttt{N} \\
\texttt{CELESTIAL\_TYPE} & \texttt{NATIVE} \\
\texttt{PROJECTION\_TYPE} & \texttt{TAN} \\
\texttt{PROJECTION\_ERR} & $0.001$ \\
\texttt{CENTER\_TYPE} & \texttt{ALL} \\
\texttt{PIXELSCALE\_TYPE} & \texttt{MEDIAN} \\
\texttt{PIXEL\_SCALE} & $0.0$ \\
\texttt{IMAGE\_SIZE} & $0$ \\
\texttt{RESAMPLE} & \texttt{Y} \\
\texttt{RESAMPLING\_TYPE} & \texttt{LANCZOS4} \\
\texttt{OVERSAMPLING} & $0$ \\
\texttt{INTERPOLATE} & \texttt{N} \\
\texttt{FSCALASTRO\_TYPE} & \texttt{NONE} \\
\texttt{FSCALE\_KEYWORD} & \texttt{NOTSCALE} \\
\texttt{FSCALE\_DEFAULT} & $1.0$ \\
\texttt{SUBTRACT\_BACK} & \texttt{Y} \\
\texttt{BACK\_TYPE} & \texttt{AUTO} \\
\texttt{BACK\_DEFAULT} & $0.0$ \\
\texttt{BACK\_SIZE} & $128$ \\
\texttt{BACK\_FILTERSIZE} & $3$ \\
\texttt{BACK\_FILTTHRESH} & $0.0$ \\
        \hline
\end{tabular}
\caption{\rm{\textsc{SWarp} parameters for matching supernova templates to single epoch science images. }}\label{tab:SWarp}
\end{table}

\begin{table}
\centering
\begin{tabular}{ll}
        \hline
Parameter & Value \\
        \hline
\texttt{useWeight} & \texttt{True} \\
\texttt{nsx} & $10$ \\
\texttt{nsy} & $20$ \\
\texttt{n} & \texttt{t} \\
\texttt{convvar} & \texttt{True} \\
\texttt{r} & $20$ \\
\texttt{rss} & $25$ \\
\texttt{kfm} & $0.999$ \\
\texttt{bgo} & $2$ \\
\texttt{rk0} & $9.01$ \\
\texttt{wk0} & $9$ \\
\texttt{wkX} & $5$ \\
\texttt{wkY} & $5$ \\
\texttt{rkx} & $5.01$ \\
\texttt{rky} & $5.01$ \\
\texttt{ng} & $1 1 6.0$ \\
\texttt{v} & $1$ \\
\texttt{mcs} & $10.0$ \\
\texttt{mds} & $5.0$ \\
\texttt{mdf} & $0.2$ \\
\texttt{mks} & $0.05$ \\
\texttt{mnor} & $5.0$ \\
\texttt{sdif} & $5.0$ \\
\texttt{serr} & $10^{-6}$ \\
\texttt{il} & $-999999$ \\
\texttt{iu} & $999999$ \\
\texttt{tl} & $-999999$ \\
\texttt{tu} & $999999$ \\
        \hline
\end{tabular}
\caption{\rm{\textsc{HOTPANTS} parameters for photometrically matching supernova template images to nightly science images and producing the resulting difference images.}}\label{tab:hotpants}
\end{table}

\begin{table}
\centering
\begin{tabular}{ll}
        \hline
Parameter & Value \\
        \hline
\texttt{DETECT\_MINAREA} & $1$ \\
\texttt{DETECT\_THRESH} & $1.2$ \\
\texttt{ANALYSIS\_THRESH} & $5.0$ \\
\texttt{FILTER} & \texttt{Y} \\
\texttt{FILTER\_NAME} & See text \\
\texttt{DEBLEND\_NTHRESH} & $32$ \\
\texttt{DEBLEND\_MINCONT} & $0.005$ \\
\texttt{CLEAN\_PARAM} & $1.0$ \\
\texttt{MASK\_TYPE} & \texttt{CORRECT} \\
\texttt{WEIGHT\_TYPE} & \texttt{MAP\_WEIGHT} \\
\texttt{WEIGHT\_GAIN} & \texttt{N} \\
\texttt{WEIGHT\_THRESH} & $1.e-10$ \\
\texttt{PHOT\_APERTURES} & $5$ \\
\texttt{PHOT\_AUTOPARAMS} & $2.5, 3.5$ \\
\texttt{PHOT\_PETROPARAMS} & $2.0, 3.5$ \\
\texttt{SATUR\_LEVEL} & $50000.0$ \\
\texttt{MAG\_ZEROPOINT} & $31.1928$ \\
\texttt{PIXEL\_SCALE} & $1.0$ \\
\texttt{SEEING\_FWHM} & $1.2$ \\
\texttt{BACK\_SIZE} & $64$ \\
\texttt{BACK\_FILTERSIZE} & $3$ \\
\texttt{BACKPHOTO\_TYPE} & \texttt{GLOBAL} \\
\texttt{CHECKIMAGE\_TYPE} & \texttt{NONE} \\
\texttt{INTERP\_MAXXLAG} & $16$ \\
\texttt{INTERP\_MAXYLAG} & $16$ \\
\texttt{INTERP\_TYPE} & \texttt{NONE} \\
        \hline
\end{tabular}
\caption{\rm{\textsc{SExtractor} parameters for detecting sources in difference images.}}\label{tab:snsex}
\end{table}

%% file: appendix/software3.tex
\section{Appendix D: Software Settings for Multi-Epoch Processing}\label{sect:software3}

Section \ref{sect:coadd} describes our Multi-Epoch Pipeline in which we coadd processed images to make deeper stacked images. We 
use \textsc{SCAMP} and the settings in Table \ref{tab:coadd-scamp} to refine the astrometric solutions 
of our input images in one mutually-consistent solution. We then combine the images into single-band 
coadds and $r + i + z$ detection images with \textsc{SWarp} 
(Table \ref{tab:coadd-swarp} and produce single-band and detection catalogs using \textsc{SExtractor} 
(Table \ref{tab:coadd-sex}). Due to the difficulties in modeling coadded PSFs, these catalogs are not used 
for our more precise analysis (see Section \ref{sect:coadd-post} for MOF description). 

\begin{table}

\centering
\begin{tabular}{ll}
        \hline
Parameter & Value \\
        \hline
\texttt{FGROUP\_RADIUS} & $3.0$\\
\texttt{ASTREF\_CATALOG} & \texttt{2MASS}\\
\texttt{ASTREF\_BAND} & \texttt{DEFAULT}\\
\texttt{ASTREFMAG\_LIMITS} & $-99$, $99$\\
\texttt{DGEOMAP\_STEP} & $2$\\
\texttt{DGEOMAP\_NNEAREST} & $21$\\
\texttt{MATCH} & \texttt{Y}\\
\texttt{MATCH\_NMAX} & $0$\\
\texttt{PIXSCALE\_MAXERR} & $1.2$\\
\texttt{POSANGLE\_MAXERR} & $0.25$\\
\texttt{POSITION\_MAXERR} & $2.0$\\
\texttt{MATCH\_RESOL} & $0$\\
\texttt{MATCH\_FLIPPED} & \texttt{N}\\
\texttt{MOSAIC\_TYPE} & \texttt{SAME\_CRVAL}\\
\texttt{FIXFOCALPLANE\_NMIN} & $1$\\
\texttt{CROSSID\_RADIUS} & $2.0$\\
\texttt{SOLVE\_ASTROM} & \texttt{Y}\\
\texttt{PROJECTION\_TYPE} & \texttt{SAME}\\
\texttt{STABILITY\_TYPE} & \texttt{INSTRUMENT}\\
\texttt{DISTORT\_GROUPS} & $1, 1$ \\
\texttt{DISTORT\_DEGREES} & $3$\\
\texttt{FOCDISTORT\_DEGREE} & $3$\\
\texttt{ASTREF\_WEIGHT} & $1.0$\\
\texttt{ASTRACCURACY\_TYPE} & \texttt{SIGMA-PIXEL}\\
\texttt{ASTR\_ACCURACY} & $0.01$\\
\texttt{ASTRCLIP\_NSIGMA} & $3.0$\\
\texttt{COMPUTE\_PARALLAXES} & \texttt{N}\\
\texttt{COMPUTE\_PROPERMOTIONS} & \texttt{N}\\
\texttt{CORRECT\_COLOURSHIFTS} & \texttt{N}\\
\texttt{INCLUDE\_ASTREFCATALOG} & \texttt{Y}\\
\texttt{ASTR\_FLAGSMASK} & $0x0000$\\
\texttt{ASTR\_IMAFLAGSMASK} & $0x000$\\
\texttt{SOLVE\_PHOTOM} & \texttt{N}\\
\texttt{MAGZERO\_OUT} & $0.0$\\
\texttt{MAGZERO\_INTERR} & $0.01$\\
\texttt{MAGZERO\_REFERR} & $0.03$\\
\texttt{PHOTCLIP\_NSIGMA} & $3.0$\\
\texttt{PHOT\_ACCURACY} & $10^{-3}$\\
\texttt{PHOT\_FLAGSMASK} & $0x00fc$\\
\texttt{PHOT\_IMAFLAGSMASK} & $0x0000$\\
\texttt{CHECKPLOT\_DEV} & \texttt{PSC}\\
\texttt{CHECKPLOT\_RES} & $0$\\
\texttt{CHECKPLOT\_ANTIALIAS} & \texttt{N}\\
\texttt{CHECKPLOT\_TYPE} & \texttt{FGROUPS,DISTORTION,}\\
                         & \texttt{ASTR\_REFERROR2D,ASTR\_REFERROR1D,}\\
                         & \texttt{ASTR\_INTERROR2D,ASTR\_INTERROR1D,}\\
                         & \texttt{ASTR\_REFSYSMAP}\\
\texttt{CHECKPLOT\_NAME} & \texttt{fgroups,distortion,}\\
                         & \texttt{astr\_referror2d,astr\_referror1d,}\\
                         & \texttt{astr\_interror1d,astr\_interror2d,}\\
                         & \texttt{astr\_refsysmap}\\
\texttt{SN\_THRESHOLDS} & $10.0$, $100.0$\\
\texttt{FWHM\_THRESHOLDS} & $1.0$, $10.0$\\
\texttt{ELLIPTICITY\_MAX} & $0.5$\\
\texttt{FLAGS\_MASK} & $0x00fd$\\
\texttt{WEIGHTFLAGS\_MASK} & $0x0000$\\
\texttt{IMAFLAGS\_MASK} & $0x0700$\\
\texttt{HEADER\_TYPE} & \texttt{NORMAL}\\
        \hline
\end{tabular}
\caption{\rm{Multi-Epoch coadd \textsc{SCAMP} settings.}}\label{tab:coadd-scamp}
\end{table}

\begin{table}

\centering
\begin{tabular}{ll}
        \hline
Parameter & Value \\
        \hline
\texttt{HEADER\_ONLY} & \texttt{N}\\
\texttt{WEIGHT\_TYPE} & \texttt{MAP\_WEIGHT}\\
\texttt{RESCALE\_WEIGHTS} & \texttt{N}\\
\texttt{COMBINE} & \texttt{Y}\\
\texttt{COMBINE\_TYPE} & \texttt{WEIGHTED}/\texttt{CHI-MEAN}\\
\texttt{CLIP\_AMPFRAC} & $0.3$\\
\texttt{CLIP\_SIGMA} & $4.0$\\
\texttt{CLIP\_WRITELOG} & \texttt{N}\\
\texttt{BLANK\_BADPIXELS} & \texttt{N}\\
\texttt{CELESTIAL\_TYPE} & \texttt{NATIVE}\\
\texttt{PROJECTION\_TYPE} & \texttt{TAN}\\
\texttt{PROJECTION\_ERR} & $0.001$\\
\texttt{CENTER\_TYPE} & \texttt{MANUAL}\\
\texttt{PIXELSCALE\_TYPE} & \texttt{MANUAL}\\
\texttt{PIXEL\_SCALE} & $0.263$\\
\texttt{IMAGE\_SIZE} & $10000$, $10000$\\
\texttt{RESAMPLE} & \texttt{Y}\\
\texttt{RESAMPLING\_TYPE} & \texttt{LANCZOS3}\\
\texttt{OVERSAMPLING} & \texttt{0}\\
\texttt{INTERPOLATE} & \texttt{N}\\
\texttt{FSCALASTRO\_TYPE} & \texttt{VARIABLE}\\
\texttt{FSCALE\_KEYWORD} & \texttt{nokey}\\
\texttt{FSCALE\_DEFAULT} & $1.0$\\
\texttt{SATLEV\_DEFAULT} & $50000.0$\\
\texttt{SUBTRACT\_BACK} & \texttt{Y}\\
\texttt{BACK\_TYPE} & \texttt{AUTO}\\
\texttt{BACK\_DEFAULT} & $0.0$\\
\texttt{BACK\_SIZE} & $256$\\
\texttt{BACK\_FILTERSIZE} & $3$\\
\texttt{BACK\_FILTTHRESH} & $0.0$\\
\texttt{COPY\_KEYWORDS} & \texttt{BUNIT,FILTER,BAND,TILENAME,TILEID}\\
        \hline
\end{tabular}
\caption{\rm{Multi-Epoch coadd \textsc{SWarp} settings. \texttt{COMBINE\_TYPE} is \texttt{WEIGHTED}
for single-band images and \texttt{CHI-MEAN} for detection images.}}\label{tab:coadd-swarp}
\end{table}

\begin{table}

\centering
\begin{tabular}{ll}
        \hline
Parameter & Value \\
        \hline
\texttt{DETECT\_TYPE} & \texttt{CCD}\\
\texttt{DETECT\_MINAREA} & $4$\\
\texttt{DETECT\_MAXAREA} & $0$\\
\texttt{THRESH\_TYPE} & \texttt{RELATIVE}\\
\texttt{DETECT\_THRESH} & 1.1\\
\texttt{ANALYSIS\_THRESH} & 1.1\\
\texttt{FILTER} & \texttt{Y}\\
\texttt{FILTER\_NAME} & \textit{See text}\\
\texttt{DEBLEND\_NTHRESH} & $32$\\
\texttt{DEBLEND\_MINCONT} & 0.001\\
\texttt{CLEAN} & \texttt{Y}\\
\texttt{CLEAN\_PARAM} & $1.0$\\
\texttt{THRESH\_TYPE} & \texttt{RELATIVE}\\
\texttt{MASK\_TYPE} & \texttt{CORRECT}\\
\texttt{WEIGHT\_TYPE} & \texttt{MAP\_WEIGHT}\\
\texttt{RESCALE\_WEIGHTS} & \texttt{N}\\
\texttt{WEIGHT\_GAIN} & \texttt{Y}\\
\texttt{FLAG\_TYPE} & \texttt{OR}\\
\texttt{PHOT\_APERTURES} & $1.85,3.70,5.55,7.41,$\\
                         & $11.11, 14.81, 18.52,$\\
                         & $22.22, 25.93, 29.63,$\\
                         & $44.44, 66.67$\\
\texttt{PHOT\_AUTOPARAMS} & $2.5, 3.5$\\
\texttt{PHOT\_PETROPARAMS} & $2.0, 3.5$\\
\texttt{PHOT\_AUTOAPERS} & $0.0, 7.41$\\
\texttt{PHOT\_FLUXFRAC} & $0.5$\\
\texttt{GAIN} & $0.0$\\
\texttt{PIXEL\_SCALE} & $0$\\
\texttt{SEEING\_FWHM} & $0$\\
\texttt{BACK\_TYPE} & \texttt{AUTO}\\
\texttt{BACK\_VALUE} & $0.0$\\
\texttt{BACK\_SIZE} & $256$\\
\texttt{BACK\_FILTERSIZE} & $3$\\
\texttt{BACKPHOTO\_TYPE} & \texttt{GLOBAL}\\
\texttt{BACKPHOTO\_THICK} & $24$\\
\texttt{BACK\_FILTTHRESH} &  $0.0$\\
\texttt{ASSOC\_DATA} & $2, 3, 4$\\
\texttt{ASSOC\_PARAMS} & $2, 3, 4$\\
\texttt{ASSOCCOORD\_TYPE} & \texttt{PIXEL}\\
\texttt{ASSOC\_RADIUS} & $2.0$\\
\texttt{ASSOC\_TYPE} & \texttt{NEAREST}\\
\texttt{ASSOCSELECT\_TYPE} & \texttt{MATCHED}\\
\texttt{VERBOSE\_TYPE} & \texttt{NORMAL}\\
\texttt{FITS\_UNSIGNED} & \texttt{N}\\
\texttt{INTERP\_MAXXLAG} & \texttt{4}\\
\texttt{INTERP\_MAXYLAG} & \texttt{4}\\
\texttt{INTERP\_TYPE} & \texttt{VAR\_ONLY}\\
\texttt{PSF\_NMAX} & \texttt{1}\\
\texttt{PATTERN\_TYPE} & \texttt{RINGS-HARMONIC}\\
        \hline
\end{tabular}
\caption{\rm{Our multiepoch coadd \textsc{SExtractor} settings.}}\label{tab:coadd-sex}
\end{table}